\documentclass[12pt]{article}
\def\baselinestretch{1.25}

\usepackage{amssymb, amsmath, amsopn, amsthm}
\usepackage{bm,amsfonts,array,calc,rotating}
\usepackage{epsfig}
\usepackage{graphicx}
\usepackage{color}

\definecolor{orange}{cmyk}{0,0.5,1,0}

\newcommand{\p}{\partial}

\newcommand{\ra}{\rightarrow}
\usepackage[T1]{fontenc}

\numberwithin{equation}{section}

\long\def\@makecaption#1#2{%
  \vskip\abovecaptionskip
  \sbox\@tempboxa{{\bf #1:} #2}%
  \ifdim \wd\@tempboxa >\hsize
    {\small\bf #1:} {\small #2}\par
  \else
    \global \@minipagefalse
    \hb@xt@\hsize{\hfil\box\@tempboxa\hfil}%
  \fi
  \vskip\belowcaptionskip}

\font\cmss=cmss10 \font\cmsss=cmss10 at 7pt
\def\IZ{\relax\ifmmode\mathchoice
{\hbox{\cmss Z\kern-.4em Z}}{\hbox{\cmss Z\kern-.4em Z}}
{\lower.9pt\hbox{\cmsss Z\kern-.4em Z}} {\lower1.2pt\hbox{\cmsss
Z\kern-.4em Z}}\else{\cmss Z\kern-.4em Z}\fi}

\def\sqr#1#2{{\vcenter{\vbox{\hrule height.#2pt
 \hbox{\vrule width.#2pt height#1pt \kern#1pt
 \vrule width.#2pt}\hrule height.#2pt}}}}

\begin{document}

\begin{flushright}
\baselineskip=12pt \normalsize
{MIFP-09-12}\\
\smallskip
\end{flushright}

\begin{center}
\Large {\textbf{A Note on Local GUT Models in F-Theory }} \\[2cm]
\normalsize Ching-Ming Chen$^{\sharp\natural}$\footnote{\tt
cchen@physics.tamu.edu} and Yu-Chieh
Chung$^{\natural}$\footnote{\tt ycchung@physics.tamu.edu}
\\[.25in]
\textit{$^{\sharp}$George P. and Cynthia W. Mitchell Institute for
Fundamental Physics, \\Texas A$\&$M University, College Station,
TX 77843, USA} \\[0.25in]
\textit{$^{\natural}$Department of Physics, Texas A$\&$M University, College Station, TX 77843, USA} \\[2.5cm]
\end{center}

\renewcommand{\baselinestretch}{1.5}
\setlength{\baselineskip}{18pt}

\begin{abstract}
We construct non-minimal GUT local models in the F-theory
configuration. The gauge group on the bulk $G_S$ is one rank
higher than the GUT gauge group. The line bundles on the curves
are non-trivial to break $G_S$ down to the GUT gauge groups.  We
demonstrate examples of $SU(5)$ GUT from $G_S=SU(6)$ and
$G_S=SO(10)$, the flipped $SU(5)$ from $G_S=SO(10)$, and the
$SO(10)$ GUT from $G_S=SO(12)$ and $G_S=E_6$. We obtain complete
GUT matter spectra and couplings, with minimum exotic matter
contents. GUT gauge group breaking to MSSM is achievable by
instanton configurations.
\end{abstract}
\vspace{4cm}


\newpage
\setcounter{page}{1}

\setcounter{footnote}{0}

\pagenumbering{arabic}

\pagestyle{plain}

\section{Introduction}

String theory is so far the most promising candidate of the
unified theory as an extension of quantum field theory and a
consistent quantum theory of gravity. It is expected to answer the
fundamental questions in physics. Many of these questions can be
explained by the extra dimensions or by the internal manifold from
the string compactification point of view. On the other hand, one
of the fundamental issues to be addressed from particle physics is
the unification of gauge couplings. The natural solution to this
question is the framework of the grand unified theory (GUT). There
are two procedures to realize GUTs in the string theory
compactification. The first is the top-down procedure in which the
full compactification is consistent with the conditions of global
geometry of extra dimensions and then the spectrum is close to GUT
after breaking some symmetries \cite{GSW}. In the bottom-up
procedure, this gauge breaking can be understood in the decoupling
limit of gravity \cite{Aldazabal:2000sa, Verlinde:2005jr},
particularly in the framework that D-branes are introduced on the
local regions within the extra dimensions in type IIB
compactification \cite{Aldazabal:2000sa, Verlinde:2005jr,
Blumenhagen:2005mu}. In this case we can neglect the effects from
the global geometry. In principle, the top-down procedure is the
more satisfactory scenario theoretically than the bottom-up
procedure. However, the later procedure is more efficient for
model building than the former one.

There is no local model in type I and heterotic string
compactifications since the matter fields live in the entire extra
dimensions.  It is possible to construct D-brane local and global
models in type IIB compacfication, however it is difficult to
engineer the ${\bf 10\,10\,5}_H$ coupling in a GUT model. This
problem can be traced to the non-realization of the exceptional
gauge groups in type IIB. In the perturbative type IIB theory, an
$SU(N)$ and an $SO(2N)$ gauge group can be realized as $N$
D-branes and $N$ D-branes along O-planes, respectively
\cite{Sen:1996vd}. The anti-symmetric representations of a GUT
come from the intersection of a stack of D-branes and its image
(as well as the orientifold), and it is not possible in this
construction to find another such intersection to finish the
Yukawa coupling without introducing exotic matter. Recently this
problem is solved in the type IIB orientifold configuration with
non-perturbative instantons corrections \cite{Blumenhagen:2008zz}
based on \cite{BlumenhagenIbanezKachru}. On the other hand, the
exceptional groups are believed to exist in the non-perturbative
regime of type IIB theory. It is well-known that the strong
coupling version of type IIB theory can be realized as F-theory
\cite{Vafa:1996xn}. Actually, those gauge groups of $ADE$-type are
naturally encoded in the geometry of the F-theory compactification
\cite{Bershadsky:1996nh, Katz:1996xe}. Thus F-theory is a natural
choice for local GUT model~building.

F-theory is a non-perturbative 12-d theory built on the type IIB
framework with an auxiliary two-torus (\cite{Vafa:1996xn}, see
\cite{Denef:2008wq} for review).  The ordinary string extra
dimensions are regarded as a base $B$ and the two-torus is
equivalent to an elliptic curve as a fiber on this base manifold.
The modulus of the elliptic curve is identified as axion-dilaton
in type IIB theory. Due to the $SL(2,Z)$ monodromy of the modulus,
F-theory is essentially non-perturbative in type IIB language. The
locations of fiber degeneracies are defined by a codimension-one
locus $\Delta$ within $B$, which also indicates the locations of
seven-branes. The fiber degeneracies lead to singularities whose
nature determines the worldvolume gauge groups of $ADE$-type on
the seven-branes \cite{Bershadsky:1996nh}.  In the strong version
of the local model, the gravity is decoupled from the gauge
theory, so we can focus on the local properties by restricting the
geometries on the submanifold $S$, which is a component of
$\Delta$ and is wrapped by seven-branes. In order to achieve that,
the volume of $S$ is required to be contractible to zero
size\footnote{There are two ways in which we could take $V_S\ra
0$. The first one is requiring $S$ to contract to a point, and the
second is requiring $S$ to contract to a curve of singularities.
See \cite{Donagi:2009ra} for the details.}, which is followed from
the condition that the anti-canonical bundle $K^{-1}_{S}$ of $S$
is ample. It implies that $S$ is a del Pezzo surface
\cite{Beasley:2008dc,donagi,Donagi:2009ra}. Given a K\"ahler
surface $S$, the maximal supersymmetric Yang-Mills theory in 8-d
admits a unique twist on $\mathbb{R}^{3,1}\times S$ which
preserves $\mathcal{N}=1$ SUSY in $\mathbb{R}^{3,1}$
\cite{Beasley:2008dc,donagi}. Matter comes from two sources, one
is from the irreducible subgroups of the bulk gauge group by
turning on nontrivial gauge bundles on $S$, and the other is from
the intersection of two del Pezzo surfaces along a codimension-two
Riemann surface $\Sigma$, which is the intersecting brane picture
in type IIB theory \cite{Katz:1996xe}. Along this curve $\Sigma$
the gauge group is enhanced and is able to be broken again by the
nontrivial gauge bundles on it. The Yukawa couplings can be
realized as couplings of either two fields from different curves
intersecting at a point and a field from the bulk, or three fields
from different curves intersecting at the same point, where the
singularity is further enhanced \cite{Beasley:2008dc,donagi}. The
generation numbers of matter on the bulk and on the curve $\Sigma$
are then determined by the dimensions of the bundle-valued
cohomology groups on $S$ and $\Sigma$, respectively
\cite{Beasley:2008dc,donagi}. One of the advantages of F-theory is
that it naturally explains the unification of the gauge couplings.

Recently some local GUT models are built in this F-theory
configuration \cite{Beasley:2008dc, donagi, Beasley:2008kw,
Heckman:2008qt, Heckman:2008rb, Font:2008id, Heckman:2008qa,
Jiang:2008yf, Blumenhagen:2008aw,Bourjaily:2009vf}, and some
progresses in global models \cite{global, tatar}. Supersymmetry
breaking is discussed in \cite{Buchbinder:2008at, Heckman:2008es,
Marsano:2008jq}, and the application to cosmology is studied in
\cite{Heckman:2008jy}. From \cite{Beasley:2008dc,
Candelas:2000nc}, the upper bound on the rank of a candidate GUT
group is six. In \cite{Beasley:2008dc,Beasley:2008kw}, the authors
consider the minimal construction by using rank four gauge group
$SU(5)$ to build $SU(5)$ GUT, and show some examples of
exotic-free models.  These models do not have the problems that a
GUT model may have, such as proton decay, doublet-triplet
splitting and so on. In this note, we shall consider non-minimal
constructions of the GUT models, namely we consider rank five and
six gauge groups to build local GUT models in F-theory.

In section $2$ of this paper, we briefly review F-theory and the
construction in \cite{Beasley:2008dc,donagi}. In section $3$, we
shall consider $SU(5)$, flipped $SU(5)$ and $SO(10)$ GUT models
from non-minimal gauge groups on $S$, and we conclude in section
$4$. In the appendices, we collect some properties of del Pezzo
surfaces and resolutions of triplet intersections for the Yukawa
couplings.

\section{F-Theory GUT Models}

The construction of local GUT models in F-theory has been analyzed
in \cite{Beasley:2008dc,donagi,Beasley:2008kw}. In this section we
shall briefly review the essential ingredient of this
construction, where the details can be found in
\cite{Beasley:2008dc,donagi,Beasley:2008kw}. Consider F-theory on
an elliptically fibered Calabi-Yau four-fold $X$ with base $B$.
Generically, the fiber degenerates on the codimension-one
reducible locus $\Delta$ within $B$. In local F-theory models, we
focus on one component $S$ of the locus $\Delta$. $S$ is a
codimension one complex surface wrapped by seven-branes and
supporting GUT models. The spirit of the bottom-up procedure leads
to the choice of $S$ being a del Pezzo surface
\cite{Beasley:2008dc,donagi,Beasley:2008kw}. To describe the
spectrum of a local model, one has to study the gauge theory of
the worldvolume on the seven-branes. As emphasized in
\cite{Beasley:2008dc,donagi}, one can start from the maximal
supersymmetric gauge theory on $\mathbb{R}^{3,1}\times
\mathbb{C}^{2}$ and then replace $\mathbb{C}^{2}$ with the
K\"ahler surface $S$. In order to make the low energy gauge theory
preserve four supercharges, the maximal supersymmetric gauge
theory on $\mathbb{R}^{3,1}\times \mathbb{C}^{2}$ should be
twisted. It is shown that there exists a unique twist preserving
${\cal N}=1$ supersymmetry in four dimensions and chiral matter
can arise from the bulk $S$ or the curve $\Sigma$
\cite{Beasley:2008dc,donagi,Beasley:2008kw}.

Let us first discuss the spectrum of the bulk fields on $S$. The
$ADE$-type singularity along $S$ is corresponding to the gauge
group $G_{S}$ on $S$ from seven-branes, and a non-trivial vector
bundle over $S$ with a structure group $H_{S}$ leads to the
unbroken gauge group $\Gamma_{S}$ in four dimensions which is the
commutant subgroup of $H_{S}$ in $G_{S}$. After compactifying on
$S$, the resulting theory is ${\cal N}=1$ supersymmetric gauge
theory with gauge group $\Gamma_{S}$ coupled to matter. The
spectrum of the bulk theory on $S$ transforms in the adjoint
representation of $G_{S}$. The decomposition of ${\rm ad} G_{S}$
into representations of $\Gamma_{S}\times H_{S}$ is
\begin{eqnarray}
{\rm ad}G_{S}=\bigoplus_{k}\rho_{k}\otimes \mathcal{R}_{k},
\end{eqnarray}
where $\rho_{k}$ and $\mathcal{R}_{k}$ are representations of
$\Gamma_S$ and $H_{S}$, respectively. The matter fields are
determined by the zero modes of the Dirac operator on $S$. It is
shown in \cite{Beasley:2008dc,donagi} that the chiral and
anti-chiral spectrum is determined by the bundle-valued cohomology
groups
\begin{equation}
H^{0}_{\bar\p}(S,R_{k}^{\vee})^{\vee}\oplus
H_{\bar\p}^{1}(S,R_{k})\oplus
H_{\bar\p}^{2}(S,R_{k}^{\vee})^{\vee}
\end{equation}
and
\begin{equation}
H_{\bar\p}^{0}(S,R_{k})\oplus
H_{\bar\p}^{1}(S,R_{k}^{\vee})^{\vee}\oplus
H_{\bar\p}^{2}(S,R_{k})
\end{equation}
respectively, where $\vee$ stands for the dual bundle and $R_{k}$
is the vector bundle on $S$ whose sections transform in the
representation $\mathcal{R}_{k}$ of the structure group $H_{S}$.
Thus, the net number of the chiral field $\rho_{k}$ and
anti-chiral field ${\rho^{\ast}_{k}}$ is given by
\begin{equation}
N_{\rho_{k}}-N_{\rho^{\ast}_{k}}=\chi(S,R_{k}^{\vee})-\chi(S,R_{k})=-\int_{S}c_{1}(R_{k})c_{1}(S).
\label{numberdiff}
\end{equation}
Moreover, by the vanishing theorem of del Pezzo surfaces
\cite{Beasley:2008dc} it shows that when $R_{k}\neq
\mathcal{O}_{S}$, then $H_{\bar\p}^{0}(S,R_{k})=0$ and
$H_{\bar\p}^{2}(S,R_{k})=0$. Thus the number of generations and
anti-generations can be calculated by
\begin{equation}
N_{\rho_{k}}=-\chi(S,R_{k})\label{numberdiff}
\end{equation}
and
\begin{equation}
N_{\rho^{\ast}_{k}}=-\chi(S,R_{k}^{\vee}),
\end{equation}
respectively.

In particular, when a gauge bundle is a line bundle $L$ with
structure group $U(1)$, according to Eq. (\ref{numberdiff}), the
chiral spectrum of $\rho_{r}$ is determined by
\begin{equation}
N_{\rho_{r}}=-\chi(S,{L}^{r})=-\Big[1+\frac{1}{2}\big(\int_{S}c_{1}({L}^{r})c_{1}(S)+
\int_{S}c_{1}({L}^{r})^2\big)\Big],\label{EulerChar}
\end{equation}
where $r$ corresponds to the $U(1)$ charges of the representations
in the group theory decomposition. In order to preserve
supersymmetry, the line bundle $L$ has to obey the BPS equation
\cite{Beasley:2008dc,donagi}
\begin{equation}
J_{S}\wedge c_{1}(L)=0,\label{BPS}
\end{equation}
where $J_{S}$ is the K\"ahler form on $S$ and its expression can
be found in the Appendix. According to Eq. (\ref{EulerChar}), by
switching on the suitable supersymmetric line bundle which
satisfies the condition $c_{1}(L)c_{1}(S)=0$, the bulk fields
$\rho_{r}$ and $\overline{\rho}_{-r}$ form a vector-like pair or
vanish, depending on the value of $c_{1}(L)^2$.

Another way to obtain chiral matter is from intersecting
seven-branes along a curve, which is a Riemann surface. Let $S$
and $S'$ be two components of the discriminant locus $\Delta$ with
gauge groups $G_{S}$ and $G_{S'}$, respectively intersecting along
a curve $\Sigma$. The gauge group on the curve $\Sigma$ will be
enhanced to $G_{\Sigma}$, where $G_{\Sigma}\supset G_{S}\times
G_{S'}$. Therefore, chiral matter appears as the bi-fundamental
representations in the decomposition of ${\rm ad}G_{\Sigma}$
\begin{equation}
{\rm ad}G_{\Sigma}={\rm ad}G_{S}\oplus {\rm ad}G_{S'}\oplus_{k}
({\cal U}_{k}\otimes {\cal U'}_{k}).
\end{equation}
As mentioned above, the presence of $H_{S}$ and $H_{S'}$ will
break $G_{S}\times G_{S'}$ to the commutant subgroup when
non-trivial gauge bundles on $S$ and $S'$ with structure groups
$H_{S}$ and $H_{S'}$ are turned on. Let
$\Gamma=\Gamma_{S}\times\Gamma_{S'}$ and $H=H_{S}\times H_{S'}$,
the decomposition of ${\cal U}\otimes {\cal U'}$ into irreducible
representation is
\begin{equation}
{\cal U}\otimes {\cal U'}={\bigoplus}_{k}(v_{k}, {\cal V}_{k}),
\end{equation}
where $v_{k}$ and ${\cal V}_{k}$ are representations of $\Gamma$
and $H$, respectively. The light chiral fermions in the
representation $v_{k}$ are determined by the zero modes of the
Dirac operator on $\Sigma$. It is shown in
\cite{Beasley:2008dc,donagi} that the net number of the chiral
field $v_{k}$ and anti-chiral field $v^{\ast}_{k}$ is given by
\begin{eqnarray}
N_{v_{k}}-N_{v^{\ast}_{k}}=\chi(\Sigma,K^{1/2}_{\Sigma}\otimes
V_{k}),
\end{eqnarray}
where $V_{k}$ is the vector bundle whose sections transform in the
representation ${\cal V}_{k}$ of the structure group $H$. In
particular, if $H_S$ and $H_{S'}$ are $U(1)$ gauge groups, the
vector bundles over $S$ and $S'$ reduce into line bundles ${L}$
and ${L}'$, respectively, then the adjoint representation ${\rm
ad}G_{\Sigma}$ will be decomposed into
\begin{equation}
{\rm ad}G_{S}\oplus {\rm ad}G_{S'}\oplus_{j}(\sigma_{j},
\sigma'_{j})_{r_{j},r'_{j}},
\end{equation}
where $r_{j}$ and $r'_{j}$ correspond to the $U(1)$ charges of the
representations in the group theory decomposition. The
bi-fundamental representation $(\sigma_{j},
\sigma'_{j})_{r_{j},r'_{j}}$ are localized on $\Sigma$
\cite{Katz:1996xe,Beasley:2008dc,donagi}. As shown in
\cite{Beasley:2008dc,donagi}, the generation number of the
representation $(\sigma_{j}, \sigma'_{j})_{r_{j},r'_{j}}$ can be
calculated by
\begin{equation}
N_{(\sigma_{j},
\sigma'_{j})_{r_{j},r'_{j}}}=h^{0}(\Sigma,K^{1/2}_{\Sigma} \otimes
{L}_{\Sigma}^{r_{j}}\otimes {L'}_{\Sigma}^{r'_{j}}),
\end{equation}
where the restrictions of line bundles to $\Sigma$ are denoted by
${L}_{\Sigma}^{r_{j}}\equiv L^{r_{j}}|_{\Sigma}$ and
${L'}_{\Sigma}^{r'_{j}}\equiv L'^{r'_{j}}|_{\Sigma}$,
respectively. It follows that the net chirality on $\Sigma$ is
given by
\begin{equation}
N_{(\sigma_{j},
\sigma'_{j})_{r_{j},r'_{j}}}-N_{\overline{(\sigma_{j},
\sigma'_{j})_{r_{j},r'_{j}}}}=c_{1}({L}_{\Sigma}^{r_{j}}\otimes
{L'}_{\Sigma}^{r'_{j}}).
\end{equation}

In addition to the analysis of the spectrum, the pattern of Yukawa
couplings is also studied \cite{Beasley:2008dc,donagi,tatar}. By
the vanishing theorem of del Pezzo surfaces
\cite{Beasley:2008dc,donagi}, Yukawa couplings can form in two
different ways. In the first type, the coupling comes from the
interaction between two fields on the curves and one field on the
bulk $S$. In the second type, all three fields are localized on
the curves which intersect at a point where the gauge group $G_p$
is further enhanced by two ranks. In the paper, we shall primarily
focus on the couplings of the second case.


\section{Model Building}

In this section we shall explore $SU(5)$, $SO(10)$ and flipped
$SU(5)$ GUT models by taking $G_{S}$ as higher rank groups. The
$SU(5)$ models from $G_S=SU(5)$ and the $SO(10)$ models from
$G_S=SO(10)$ have been discussed in \cite{Beasley:2008dc,donagi,
Beasley:2008kw}. In these models, the restriction of line bundles
on the bulk to the matter curves are required to be trivial to
maintain the GUT fermion spectrum, while they are nontrivial on
the curves for Higgs fields to explain the phenomenology of
doublet-triplet splitting when GUT breaks to the Minimum
Supersymmetric Standard Model (MSSM). The curve self-intersection
mechanism makes it possible to explain the rank three quark and
lepton mass matrices from the Yukawa couplings. The bulk line
bundle can be nontrivial on the matter curves, which is useful in
discussing a flipped $SU(5)$ model \cite{Jiang:2008yf}, and a rich
SM Yukawa mass structure \cite{Font:2008id}.

We shall mainly focus on the cases that the gauge groups on $S$
have higher ranks than the GUT gauge groups, so the bulk line
bundles will be nontrivial on all the curves to obtain GUT
spectra. There is no GUT adjoint representation on a del Pezzo
surface, but it is still possible to break the GUT gauge groups to
the Standard Model (SM) gauge group by introducing non-Abelian
instanton configurations on the bulk \cite{Beasley:2008kw}.  For
the maximum degrees of freedom of model building, the del Pezzo
surfaces in the following models are all $dP_8$.

\subsection{$SU(5)$ GUT}

\subsubsection{$G_{S}=SU(6)$}
\label{secsu5su6}

Consider seven-branes wrapping on a del Pezzo surface $S=dP_8$
with $G_{S}=SU(6)$. From Eq.~(\ref{EulerChar}), the bulk field
${\rho_{r}}$ is determined by the bundle-valued Euler
characteristic $\chi(S,{L}^{r})$ where $r$ is the $U(1)$ charge in
the group theory decomposition. According to the property of the
Chern class, $c_{n}(L^{-r})=(-1)^{n}c_{n}(L^{r})$, where $L^{-r}$
is the dual bundle of $L^{r}$. In particular, when $n=1$ we obtain
$c_{1}(L^{-r})=-c_{1}(L^{r})$, and it turns out that
$N_{\rho_{r}}-N_{\overline{\rho}_{-r}}=-r\int c_{1}(L)c_{1}(S)$.
If $N_{\rho_{r}}\neq 0$, it implies that the bulk fields
$\rho_{r}$ and $\overline{\rho}_{-r}$ form a vector-like pair if
\begin{equation}
c_{1}(L)c_{1}(S)=0, \label{CClassLS}
\end{equation}
for example $\displaystyle
L=\mathcal{O}_{S}(\sum_{m=1}^{2l}(-1)^{m+1}E_{i_{m}}), \;l\leq 4$,
where all indices are distinct. It is easy to see that it solves
Eq.~(\ref{CClassLS}) and the BPS equation (\ref{BPS}) by choosing
suitable polarization of $J_S$, for example
$J_{S}=AH-\sum^{8}_{i=1}E_{i},\;A\gg 1$. If $L$ is a line bundle
satisfying $\chi(S,{L}^{r})=\chi(S,{L}^{-r})=0$, then
$N_{\rho_{r}}=N_{\overline{\rho}_{-r}}=0$. In other words, no
chiral field lives on the bulk. In this case, it is not difficult
to find that $L=\mathcal{O}_{S}(E_{i}-E_{j})^{1/r},\;i\neq j$,
which is a well-defined fractional line bundle\footnote{It is not
the only solution, for example, it could be
$L=\mathcal{O}_{S}(\sum_{m=1}^{8}(-1)^{m+1}E_{m})^{1/2r}$.
However, $L=\mathcal{O}_{S}(E_{i}-E_{j})^{1/r},\;i\neq j$, is the
only solution that $c_1(L^{r})\in H_2(S,\mathbb{Z}$).} due to the
fact that $c_{1}(L^{r})$ is a integer class
\cite{Beasley:2008dc,donagi,Beasley:2008kw}.

In this model where $G_{S}=SU(6)$, the possible breaking patterns
on the local curve by $U(1)$ line bundle from $S'$ and by $U(1)_S$
line bundle on the bulk are \cite{Slansky:1981yr}:

\begin{equation}
\begin{array}{c@{}c@{}l@{}c@{}l}
SU(7) &~\rightarrow~& SU(6)_S\times U(1) &~\rightarrow~& SU(5)\times U(1)\times U(1)_S\\
48 &~\rightarrow~& 35_0+1_0&~\rightarrow~&
24_{0,0}+1_{0,0}+{5_{0,6}}+\bar 5_{0,-6}+1_{0,0} \\ & &
+6_{-7}+\bar 6_{7}& &+5_{-7,1}+1_{-7,-5}+\bar 5_{7,-1}+1_{7,5}
\end{array}
\end{equation}
\begin{equation}
\begin{array}{c@{}c@{}l@{}c@{}l}
SO(12) &~\rightarrow~& SU(6)_S\times U(1) &~\rightarrow~&
SU(5)\times U(1)\times U(1)_S\\ 66 &~\rightarrow~&
35_0+1_0&~\rightarrow~& 24_{0,0}+1_{0,0}+{5_{0,6}}+\bar
5_{0,-6}+1_{0,0}\\ & &+15_2+\overline{15}_{-2} &
&+10_{2,2}+5_{2,-4}+\overline{10}_{-2,-2}+\bar 5_{-2,4}
\end{array}
\end{equation}
\begin{equation}
\begin{array}{c@{}c@{}l@{}c@{}l}
E_{6} &~\rightarrow~& SU(6)_S\times U(1) &~\rightarrow~& SU(5)\times U(1)\times U(1)_S\\
78 &~\rightarrow~& 35_0+1_0+{\it 1_{\pm2}}&~\rightarrow~&
24_{0,0}+2\times 1_{0,0}+5_{0,6}+ \bar{5}_{0,-6} +{\it 1_{\pm2,0}}\\
& & +20_1+{\it {20}_{-1}} & &+10_{1,-3}+\overline{10}_{1,3}+\it
10_{-1,-3}+\overline{10}_{-1,3}
\end{array}
\end{equation}

We shall consider the supersymmetric line bundle
$L=\mathcal{O}_{S}(E_1-E_2)^{1/6}$ so that there is no chiral
field on the bulk, $i.e.$ $N_{{\bf 5}_{6}}=N_{{\bf\bar
5}_{-6}}=0$. Therefore, there is no Yukawa coupling of
$\Sigma\Sigma S$ type, such as ${\bf 10}_{-1,-3}{\bf
10}_{-1,-3}{\bf 5}_{0,6}$, ${\bf 10}_{2,2} \bar{\bf
5}_{-2,4}\bar{\bf 5}_{0,-6}$ and their complex conjugates. The
first $U(1)$ charge of each representation is from $S'$ and the
second is from the bulk. Since the bulk line bundle is not trivial
in our discussion, the $U(1)_S$ charges should be conserved in
each Yukawa coupling.

Now let us turn to the chiral spectra from the curves. The same
representation can come from alternate breaking patterns giving
varied charges.  The difference from the cases in
\cite{Beasley:2008dc, Beasley:2008kw} is that the restriction of
bulk fluxes to the matter curves are nontrivial here. Therefore we
have to choose proper representations from the curves that
intersect at a double enhanced point forming the corresponding
Yukawa coupling. One possible choice of such $SU(5)$ model from
$G_S=SU(6)$ in terms of the matter representations on the curves
is:
\begin{eqnarray}
\mathcal{W}&\supset & {\bf 10}_{2,2} {\bf 10}_{2,2} {\bf 5}_{2,-4}
+{\bf 10}_{2,2} \bar{\bf 5}_{7,-1} \bar{\bf 5}_{7,-1}+\cdots.
\end{eqnarray}
The corresponding Yukawa coupling patterns on the double enhanced
points of $\bf 10\, 10\,5$ and ${\bf 10}\,{\bar{\bf 5}}\,{\bar{\bf
5}}$ can be found in Eq. (\ref{su610105}) and Eq.
(\ref{su6105-5-}), respectively.

In what follows, we engineer the minimal spectrum by introducing
suitable supersymmetric line bundles. Let $L$ and $L'$ be the line
bundles over $S$ and $S'$ respectively, and consider $\Sigma$ to
be a curve of genus zero. Let
$L_{\Sigma}=\mathcal{O}_{\Sigma}(a_{\Sigma})$ and
$L'_{\Sigma}=\mathcal{O}_{\Sigma}(b_{\Sigma})$ be the line bundles
restricted to the curve $\Sigma$.  The parameters $a_{\Sigma}$ and
$b_{\Sigma}$ from the line bundles $L$ and $L'$ need to be fixed
by the constraints from the matter spectrum, and there could be
more than two conditions from these constraints resulting in the
existence of exotic matter.

According to \cite{tatar}, it is not necessary to use the
self-intersecting mechanism in
\cite{Beasley:2008dc,Beasley:2008kw} to obtain the codimension
three Yukawa coupling ${\bf 10\,10\,5}_H$, and one can instead
simply engineer two intersecting curves supporting ${\bf
10}_{2,2}$ and ${\bf 5}_{2,-4}$ to get a rank one coupling. We
will follow the latter to construct the Yukawa coupling.

The three generations are from the curve $\Sigma_{M}^{1}$ with the
enhanced group $G_{\Sigma_{M}^{1}}=SO(12)$. Let the line bundles
on this curve be
$L_{\Sigma_{M}^{1}}=\mathcal{O}_{\Sigma_{M}^{1}}(a_{M}^{1})$ and
$L'_{\Sigma_{M}^{1}}=\mathcal{O}_{\Sigma_{M}^{1}}(b_{M}^{1})$. It
is required to obtain the desired field content that
\[h^{0}(\Sigma_{M}^{1},K^{1/2}_{\Sigma_{M}^{1}}\otimes\mathcal{O}_{\Sigma_{M}^{1}}(2a_{M}^{1})
\otimes\mathcal{O}_{\Sigma_{M}^{1}}(2b_{M}^{1}))=3,\]
\[h^{0}(\Sigma_{M}^{1},K^{1/2}_{\Sigma_{M}^{1}}\otimes\mathcal{O}_{\Sigma_{M}^{1}}(-2a_{M}^{1})
\otimes\mathcal{O}_{\Sigma_{M}^{1}}(-2b_{M}^{1}))=0,\]
\[h^{0}(\Sigma_{M}^{1},K^{1/2}_{\Sigma_{M}^{1}}\otimes\mathcal{O}_{\Sigma_{M}^{1}}(-4a_{M}^{1})
\otimes\mathcal{O}_{\Sigma_{M}^{1}}(2b_{M}^{1}))=0,\]
\[h^{0}(\Sigma_{M}^{1},K^{1/2}_{\Sigma_{M}^{1}}\otimes\mathcal{O}_{\Sigma_{M}^{1}}(4a_{M}^{1})
\otimes\mathcal{O}_{\Sigma_{M}^{1}}(-2b_{M}^{1}))=0.\] It is easy
to find that the unique solution is $a_{M}^{1}=\frac{1}{2}$ and
$b_{M}^{1}=1$, so there exist
\[3\times {\bf 10}_{2,2}\]
localized on the curve $\Sigma_{M}^{1}$.

Let the matter multiple $\bf \bar 5$ be from the curve
$\Sigma_{M}^{2}$. We choose this curve to be genus zero with the
enhanced group $G_{\Sigma_{M}^{2}}=SU(7)$ and the line bundles on
$\Sigma_{M}^{2}$ to be
$L_{\Sigma_{M}^{2}}=\mathcal{O}_{\Sigma_{M}^{1}}(a_{M}^{2})$ and
$L'_{\Sigma_{M}^{2}}=\mathcal{O}_{\Sigma_{M}^{1}}(b_{M}^{2})$. In
this case, we obtain the unique solution $a_{M}^{2}=-\frac{1}{2}$
and $b_{M}^{2}=\frac{5}{14}$. The resulting field content is
\[3\times\bar{\bf 5}_{7,-1}.\]

Let the up-type Higgs multiplet be from the curve
$\Sigma_{H}^{1}$. Then we choose it also a genus zero curve with
the enhanced group $G_{\Sigma_{H}^{1}}=SO(12)$ and the line
bundles on $\Sigma_{H}^{1}$ as
$L_{\Sigma_{H}^{1}}=\mathcal{O}_{\Sigma_{H}^{1}}(a_{H}^{1})$ and
$L'_{\Sigma_{H}^{1}}=\mathcal{O}_{\Sigma_{H}^{1}}(b_{H}^{1})$. The
unique solution is $a_{H}^{1}=-\frac{1}{6}$ and
$b_{H}^{1}=\frac{1}{6}$, so the field content is
\[1\times {\bf 5}_{2,-4}.\]
Similarly, for the down-type Higgs multiplet on $\Sigma_{H}^{2}$,
we again take it as a genus zero curve with the enhanced group
$G_{\Sigma_{H}^{2}}=SU(7)$ and the line bundles on
$\Sigma_{H}^{2}$ are
$L_{\Sigma_{H}^{2}}=\mathcal{O}_{\Sigma_{H}^{2}}(a_{H}^{2})$ and
$L'_{\Sigma_{H}^{2}}=\mathcal{O}_{\Sigma_{H}^{2}}(b_{H}^{2})$. In
this case, we obtain the unique solution $a_{H}^{2}=-\frac{1}{6}$
and $b_{H}^{2}=\frac{5}{42}$ and the field content is
\[1\times \bar{\bf 5}_{7,-1}.\]

After determining the line bundles, we look for the suitable
curves to support these bundles.  In our construction we require
all curves effective and genus zero. Of course it is possible to
choose the curves with higher genus, such as a genus one curve
with non-effective divisors. However, there will exist vector-like
Higgs fields on these curves, which may result in the problem of
rapid proton decay \cite{Beasley:2008kw}. Therefore, we only
consider curves of genus zero and separate up-type and down-type
Higgs fields on different curves.

We summarize the spectrum and the homology classes of the curves
of this model in Table {\ref{SU(5)SU(6)B}}.

\begin{table}[h]
\begin{center}
\renewcommand{\arraystretch}{1.25}
\begin{tabular}{|c|c|c|c|c|c|c|c|} \hline

Multiplet & Curve & Class & $g_{\Sigma}$ & $L_{\Sigma}$ & $L'_{\Sigma}$ \\
\hline \hline

$3\times {\bf 10}_{2,2}$ & $\Sigma_{M}^{1}$ & $4H+2E_2-E_1$ & 0 &
$\mathcal{O}_{\Sigma_{M}^{1}}(1)^{1/2}$ & $\mathcal{O}_{\Sigma_{M}^{1}}(1)$ \\
\hline

$3\times\bar{\bf 5}_{7,-1}$ & $\Sigma_{M}^{2}$ & $5H+3E_1-E_6$ & 0
&
$\mathcal{O}_{\Sigma_{M}^{2}}(-1)^{1/2}$ & $\mathcal{O}_{\Sigma_{M}^{2}}(1)^{5/14}$ \\
\hline

$1\times {\bf 5}_{2,-4}$ & $\Sigma_{H}^{1}$ & $3H+E_1-E_3$ & 0 &
$\mathcal{O}_{\Sigma_{H}^{1}}(-1)^{1/6}$ & $\mathcal{O}_{\Sigma_{H}^{1}}(1)^{1/6}$ \\
\hline

$1\times\bar{\bf 5}_{7,-1}$ & $\Sigma_{H}^{2}$ & $H-E_2-E_3$ & 0 &
$\mathcal{O}_{\Sigma_{H}^{2}}(-1)^{1/6}$ & $\mathcal{O}_{\Sigma_{H}^{2}}(1)^{5/42}$ \\
\hline
\end{tabular}
\caption{An $SU(5)$ GUT model from $G_S=SU(6)$, where
$L=\mathcal{O}_{S}(E_{1}-E_{2})^{1/6}$.} \label{SU(5)SU(6)B}
\end{center}
\end{table}


\subsubsection{$G_{S}=SO(10)$}
\label{secsu5so10}

Consider a $G_S=SO(10)$ model with nontrivial line bundles on all
the curves, so $SO(10)$ is broken down to $SU(5)\times U(1)_S$ on
the bulk. Like the previous case, we choose a supersymmetric line
bundle $L=\mathcal{O}_{S}(E_{1}-E_{2})^{1/4}$ on $S$ such that the
chiral matter fields on the bulk vanish, $i.e. $ $N_{{\bf
10}_{4}}=N_{\overline{{\bf 10}}_{-4}}=0$.  The Yukawa couplings of
$\Sigma$$\Sigma$S type such as ${\bf 10}_{0,4}\bar {\bf
5}_{2,-2}\bar {\bf 5}_{-2,-2}$ and ${\bf 10}_{0,4}{\bf
10}_{-3,-1}{\bf 5}_{3,-3}$ and their complex conjugates are
vanishing. We shall only consider the Yukawa couplings of
$\Sigma$$\Sigma\Sigma$ type where chiral fields are from local
curves $\Sigma$\,s in the following example.

The breaking chains and matter content from the enhanced adjoints
of the curves are
\begin{equation}
\begin{array}{c@{}c@{}l@{}c@{}l}
SO(12) &~\rightarrow~& SO(10)_S\times U(1) &~\rightarrow~& SU(5)\times U(1)\times U(1)_S\\
66 &~\rightarrow~& 45_0+1_0&~\rightarrow~&
24_{0,0}+1_{0,0}+10_{0,4}+ \overline{10}_{0,-4}+1_{0,0}\\
& &+10_2+\overline{10}_{-2} & &+5_{2,2}+{\bar5_{2,-2}}+{\bar
5_{-2,-2}}+5_{-2,2}
\end{array}
\end{equation}
\begin{equation}
\begin{array}{c@{}c@{}l@{}c@{}l}
E_6 &~\rightarrow~ & SO(10)_S\times U(1) &~\rightarrow~& SU(5)\times U(1)\times U(1)_S\\
78 &~\rightarrow~ & 45_0+1_0&~\rightarrow~&
24_{0,0}+1_{0,0}+{10_{0,4}}+ \overline{10}_{0,-4}+1_{0,0}\\
& & +16_{-3}+\overline{16}_{3} & &+(10_{-3,-1}+{\bar 5}_{-3,3}+1_{-3,-5}+c.c.)\\
\end{array}
\end{equation}

Let us turn to the spectrum from the curves. Again, since the bulk
line bundle is nontrivial in our discussion, the $U(1)_S$ charges
of the fields localized on the curves should be conserved in each
Yukawa coupling. The superpotential is:
\begin{eqnarray}
\mathcal{W}&\supset & {\bf 10}_{-3,-1} {\bf 10}_{-3,-1} {\bf
5}_{-2,2}+{\bf 10}_{-3,-1} \bar{\bf 5}_{-3,3} \bar{\bf
5}_{2,-2}+\cdots.
\end{eqnarray}
The corresponding Yukawa coupling patterns on the double enhanced
points of $\bf 10\, 10\,5$ and ${\bf 10}\,{\bar{\bf 5}}\,{\bar{\bf
5}}$ can be found in Eq. (\ref{so1010105}) and Eq.
(\ref{so10105-5-}), respectively.

To obtain the spectrum, first we choose the genus zero curve
$\Sigma_{M}^{1}$ with $G_{\Sigma_{M}^{1}}=E_{6}$ and let
$L_{\Sigma_{M}^{1}}=\mathcal{O}_{\Sigma_{M}^{1}}(d_{M}^{1})$ and
$L'_{\Sigma_{M}^{1}}=\mathcal{O}_{\Sigma_{M}^{1}}(e_{M}^{1})$. In
order to get the desired field content, it is required that
\[h^{0}(\Sigma_{M}^{1},K^{1/2}_{\Sigma_{M}^{1}}\otimes\mathcal{O}_{\Sigma_{M}^{1}}(-d_{M}^{1})
\otimes\mathcal{O}_{\Sigma_{M}^{1}}(-3e_{M}^{1}))=3,\]
\[h^{0}(\Sigma_{M}^{1},K^{1/2}_{\Sigma_{M}^{1}}\otimes\mathcal{O}_{\Sigma_{M}^{1}}(d_{M}^{1})
\otimes\mathcal{O}_{\Sigma_{M}^{1}}(3e_{M}^{1}))=0,\]
\[h^{0}(\Sigma_{M}^{1},K^{1/2}_{\Sigma_{M}^{1}}\otimes\mathcal{O}_{\Sigma_{M}^{1}}(3d_{M}^{1})
\otimes\mathcal{O}_{\Sigma_{M}^{1}}(-3e_{M}^{1}))=0,\]
\[h^{0}(\Sigma_{M}^{1},K^{1/2}_{\Sigma_{M}^{1}}\otimes\mathcal{O}_{\Sigma_{M}^{1}}(-3d_{M}^{1})
\otimes\mathcal{O}_{\Sigma_{M}^{1}}(3e_{M}^{1}))=0,\]
\[h^{0}(\Sigma_{M}^{1},K^{1/2}_{\Sigma_{M}^{1}}\otimes\mathcal{O}_{\Sigma_{M}^{1}}(-5d_{M}^{1})
\otimes\mathcal{O}_{\Sigma_{M}^{1}}(-3e_{M}^{1}))=0,\]
\[h^{0}(\Sigma_{M}^{1},K^{1/2}_{\Sigma_{M}^{1}}\otimes\mathcal{O}_{\Sigma_{M}^{1}}(5d_{M}^{1})
\otimes\mathcal{O}_{\Sigma_{M}^{1}}(3e_{M}^{1}))=0.\] It is easy
to see no solution satisfies all conditions, which means that
there exists exotic matter. We choose $d_{M}^{1}=-\frac{3}{4}$ and
$e_{M}^{1}=-\frac{3}{4}$, then the field content includes exotic
singlets:
\[3\times {\bf 10}_{-3,-1},\;\;\;6\times {\bf 1}_{-3,-5}.\]

For $\Sigma_{M}^{2}$, we take it as a genus zero curve with
$G_{\Sigma}=E_{6}$ and let the line bundles be
$L_{\Sigma_{M}^{2}}=\mathcal{O}_{\Sigma_{M}^{2}}(d_{M}^{2})$ and
$L'_{\Sigma_{M}^{2}}=\mathcal{O}_{\Sigma_{M}^{2}}(e_{M}^{2})$.
Again, no solution satisfies all the conditions, which means that
there exists exotic matter. We choose $d_{M}^{2}=\frac{3}{4}$ and
$e_{M}^{2}=-\frac{1}{4}$ so then the field content is
\[3\times \overline{\bf 5}_{-3,3},\;\;\;3\times {\bf 1}_{3,5}.\]

We choose $\Sigma_{H}^{1}$ to be a genus zero curve with
$G_{\Sigma_{H}^{1}}=SO(12)$ and let the line bundles be
$L_{\Sigma_{H}^{1}}=\mathcal{O}_{\Sigma_{H}^{1}}(d_{H}^{1})$ and
$L'_{\Sigma_{H}^{1}}=\mathcal{O}_{\Sigma_{H}^{1}}(e_{H}^{1})$. The
unique solution is $d_{H}^{1}=\frac{1}{4}$ and
$e_{H}^{1}=-\frac{1}{4}$. The resulting field content is
\[1\times {\bf 5}_{-2,2}.\]

We choose $\Sigma_{H}^{2}$ to be a genus zero curve with
$G_{\Sigma_{H}^{2}}=SO(12)$ and let the line bundles be
$L_{\Sigma_{H}^{2}}=\mathcal{O}_{\Sigma_{H}^{2}}(d_{H}^{2})$ and
$L'_{\Sigma_{H}^{2}}=\mathcal{O}_{\Sigma_{H}^{2}}(e_{H}^{2})$. The
solution is $d_{H}^{2}=-\frac{1}{4}$ and $e_{H}^{2}=\frac{1}{4}$,
thus the resulting field content is
\[1\times {\bf\bar 5}_{2,-2}.\]
We summarize the result in Table \ref{SU5SO10}.

\begin{table}[h]
\begin{center}
\renewcommand{\arraystretch}{1.25}
\begin{tabular}{|c|c|c|c|c|c|c|} \hline

Multiplet & Curve & Class & $g_{\Sigma}$ & $L_{\Sigma}$ & $L'_{\Sigma}$ \\
\hline \hline

$3\times {\bf 10}_{-3,-1}$ & $\Sigma_{M}^{1}$ & $4H+2E_1-E_2$ & 0
&
$\mathcal{O}_{\Sigma_{M}^{1}}(-1)^{3/4}$ & $\mathcal{O}_{\Sigma_{M}^{1}}(-1)^{3/4}$ \\
\hline

$3\times\bar {\bf 5}_{-3,3}$ & $\Sigma_{M}^{2}$ & $5H+3E_2-E_5$ &
0 &
$\mathcal{O}_{\Sigma_{M}^{2}}(1)^{3/4}$ & $\mathcal{O}_{\Sigma_{M}^{2}}(-1)^{1/4}$ \\
\hline

$1\times {\bf 5}_{-2,2}$ & $\Sigma_{H}^{1}$ & $3H+E_3-E_1$ & 0 &
$\mathcal{O}_{\Sigma_{h}^{1}}(1)^{1/4}$ & $\mathcal{O}_{\Sigma_{h}^{1}}(-1)^{1/4}$ \\
\hline

$1\times\bar{\bf 5}_{2,-2}$ & $\Sigma_{H}^{2}$ & $H-E_2-E_3$
& 0 & $\mathcal{O}_{\Sigma_{h}^{2}}(-1)^{1/4}$ & $\mathcal{O}_{\Sigma_{h}^{2}}(1)^{1/4}$ \\
\hline

\end{tabular}
\caption{An $SU(5)$ GUT model from $G_S=SO(10)$, where
$L=\mathcal{O}_{S}(E_{1}-E_{2})^{1/4}$.}\label{SU5SO10}
\end{center}
\end{table}


In the first example with $G_S=SU(6)$, the flux is nontrivial in
order to break the bulk gauge group into the desired $SU(5)$ gauge
group. We choose the case that all matter fields come from the
curves without exotic fields. We avoid the possibilities of
up-type and down-type Higgs fields coming from the bulk or from
the same curve that will cause rapid proton decay by the induced
quartic terms in the superpotential. The $U(1)_S$ charges are
consistent in the fermion mass Yukawa couplings.

In the second example with $G_S=SO(10)$, the flux is nontrivial as
well in order to break the bulk gauge group into the desired
$SU(5)$ gauge group. All matter fields are from the curves without
exotic fields on the bulk. The $U(1)_S$ charges are consistent in
the Yukawa couplings and it explains that an $SU(5)$ GUT is
descended from the $SO(10)$ unified gauge group.

\subsubsection{Split gauge bundle}

The Standard Model (SM) gauge group is two ranks lower than $G_S$,
therefore in principle, if we want to break $G_S$ to $SU(3)\times
SU(2)\times U(1)_Y$ it is possible to introduce an instanton
configuration to break $G_S$ \cite{Beasley:2008kw}. This instanton
can be a $SU(2)$ or $U(1)\times U(1)$ gauge group.  In the models
discussed above, the $U(1)_S$ is a substructure of $U(1)\times
U(1)$, and the additional $U(1)_{\tilde S}$ can be utilized on the
bulk to break the $SU(5)$ GUT to SM. $U(1)_Y$ which can be the
linear combination of these $U(1)$s. In this case, the
$U(1)_{\tilde S}$ charges are consistent with the $U(1)_Y$
charges. There is also a possibility to solve the doublet-triblet
problem from controlling the Higgs multiplets by this
$U(1)_{\tilde S}$ gauge group.  In what follows we demonstrate an
example that how this Abelian gauge bundle breaks the $SU(5)$ GUT
group on the bulk.

Consider $V$ to be a split vector bundle of rank two over $S$.
Write $V=L_1\oplus L_2$, where $L_i,\;i=1,2$ are nontrivial line
bundles. In order to solve the BPS equation $(\ref{BPS})$, the
line bundles are required to be supersymmetric, in other words,
$J_S\wedge c_{1}(L_1)=J_S\wedge c_{1}(L_2)=0$. To be more
concrete, let $V=\mathcal{O}_{S}(E_i-E_j)\oplus
\mathcal{O}_{S}(E_j-E_i)^{1/6},\;i\neq j$, it is easy to check
that it solves BPS equation. In this case, the structure group is
$U(1)_{\tilde S}\times U(1)_{S}$. Therefore, by switching on the
gauge bundle $V$, $G_S=SU(6)$ can be broken into $SU(3)\times
SU(2)\times U(1)_{\tilde S}\times U(1)_{S}$. The breaking pattern
is as follows
\begin{equation}
\begin{array}{c@{}c@{}l@{}c@{}l}
SU(6) &~\rightarrow~& SU(3)\times SU(2)\times U(1)_{\tilde S}\times U(1)_{S} \\
35 &~\rightarrow~& (8,1)_{0,0}+(1,3)_{0,0}+(3,2)_{-5,0}+(\bar
3,2)_{5,0}+(1,1)_{0,0}\\&
&+(1,1)_{0,0}+(1,2)_{3,6}+(3,1)_{-2,6}+(1,\bar 2)_{-3,-6}+(\bar
3,1)_{2,-6}
\end{array}
\end{equation}
It turns out that in this case, all fields on the bulk form
vector-like pairs. The spectrum on the bulk is then given by
\begin{equation} \left\{\begin{array}{l}
N_{(3,2)_{-5,0}}=N_{(\bar
3,2)_{5,0}}=24\\N_{(1,2)_{3,6}}=N_{(1,\bar 2)_{-3,-6}}=3 \\
N_{(\bar 3,1)_{2,-6}}=N_{(3,1)_{-2,6}}=8  \\
\end{array}   \right.
\end{equation}
Of course this is not the only choice for the split gauge bundle
of rank two over $S$. The detailed configuration and the spectrum
of the chiral fields from curves will be presented elsewhere
\cite{Chung:2009ib}.

The self-intersection mechanism of the $\bf 10$ curve in the $\bf
10\,10\,5$ coupling is not the only way to obtain higher rank
Yukawa mass matrices. It has been shown in \cite{Font:2008id} that
a generalization of the conditions on the $U(1)_{B-L}$ flux with
the chiral fermions from two different curves can take the work.
With the introduction of this additional $U(1)$, the generation
numbers of MSSM fields in the $\bf 10$ and $\bf 5$ representations
of $SU(5)$ can be controlled to achieve a richer structure of the
fermion mass matrices.

\subsection{Flipped $SU(5)$ GUT}

In a flipped $SU(5)\times U(1)_X$ \cite{smbarr,dimitri,AEHN-0}
unified model, the electric charge generator is only partially
embedded in SU(5). In other words, the photon is shared between
$SU(5)$ and $U(1)_X$. The SM fermions plus the right-handed
neutrino states reside within the representations $\bar{\bf 5}$,
$\bf 10$, and $\bf 1$ of $SU(5)$, which are collectively
equivalent to a spinor $\bf 16$ of $SO(10)$. The quark and lepton
assignments are flipped by $u^c_L \leftrightarrow d^c_L$ and
$\mu^c_L \leftrightarrow e^c_L$ relative to a conventional $SU(5)$
GUT embedding. Since $\bf 10$ contains a neutral component
$\nu^c_L$, we can spontaneously break the GUT gauge symmetry by
using a pair of ${\bf 10}_H$ and $\overline{\bf 10}_H$ of
superheavy Higgs where the neutral components receive a large VEV.
The spontaneous breaking of electroweak gauge symmetry is
generated by the Higgs doublets embedded in the Higgs pentaplet
${\bf 5}_h$. It then has a natural solution to the doublet-triplet
splitting problem through the trilinear coupling of the Higgs
fields ${\bf 10}_H{\bf 10}_H{\bf 5}_h$. The generic superpotential
$\mathcal{W}$ is
\begin{equation}
\mathcal{W}\supset {\bf 10\,10\,5}_h + {\bf 10}\,\bar{\bf
5}\,\bar{\bf 5}_h + \bar{\bf 5}\,{\bf 1}\, {\bf 5}_h + {\bf
10}~\overline{\bf 10}_H {\bf 1}_{\phi} + {\bf 10}_H {\bf 10}_H
{\bf 5}_h + \overline{\bf 10}_H \overline{\bf 10}_H \bar {\bf
5}_h+\cdots. \label{superFSU5}
\end{equation}

\subsubsection{$G_{S}=SU(6)$}

Since the flipped $SU(5)$ model has a similar fermion spectrum as
the $SU(5)$ model, and there are limited options for the matter
from the curves, we may make the $SU(5)\times U(1)_X$ model from
$G_S=SU(6)$ based on the setup of the previous section
(\ref{secsu5su6}) with additional fields such as the singlet ${\bf
1}_{M}$ and the GUT Higgs ${\bf 10}_{H}$, $\overline{\bf 10}_{H}$.
One possible choice for the Yukawa couplings is:
\begin{eqnarray}
\mathcal{W}\supset  {\bf 10}_{2,2} {\bf 10}_{2,2} {\bf 5}_{2,-4}
+{\bf 10}_{2,2} \bar{\bf 5}_{7,-1} \bar{\bf 5}_{7,-1} +\bar{\bf
5}_{7,-1}{\bf 5}_{2,-4}{\bf 1}_{7,5}+\cdots.
\end{eqnarray}

The construction is similar to the $SU(5)$ model from $G_S=SU(6)$
in the previous section, and we need the additional matter singlet
and the superheavy Higgs pairs. We choose $\Sigma_{M}^{3}$ to be a
genus zero curve with $G_{\Sigma_{M}^{3}}=SU(7)$ and let the line
bundles be
$L_{\Sigma_{M}^{3}}=\mathcal{O}_{\Sigma_{M}^{3}}({\tilde
a}^{3}_{M})$ and
$L'_{\Sigma_{M}^{3}}=\mathcal{O}_{\Sigma_{M}^{3}}({\tilde
b}^{3}_{M})$. The unique solution is ${\tilde
a}^{3}_{M}=\frac{1}{2}$ and ${\tilde b}^{3}_{M}=\frac{1}{14}$ and
the resulting field content is \[3\times {\bf 1}_{7,5}.\]

We choose $\Sigma_{H}^{1}$ to be a genus zero curve with
$G_{\Sigma_{H}^{1}}=SO(12)$. Let
$L_{\Sigma_{H}^{1}}=\mathcal{O}_{\Sigma_{H}^{1}}({\tilde
a}^{1}_{H})$ and
$L'_{\Sigma_{H}^{1}}=\mathcal{O}_{\Sigma_{H}^{1}}({\tilde
b}^{1}_{H})$, the unique solution is ${\tilde
a}^{1}_{H}=\frac{1}{6}$ and ${\tilde b}^{1}_{H}=\frac{1}{3}$ and
the resulting field content is
\[1\times {\bf 10}_{2,2}.\]
Similarly, for $\Sigma_{H}^{2}$, we make it genus zero. The
resulting field content is \[1\times \overline{\bf 10}_{-2,-2}.\]
We summarize the pinched model in Table \ref{SU6FSU5}.

\begin{table}[h]
\begin{center}
\renewcommand{\arraystretch}{1.25}
\begin{tabular}{|c|c|c|c|c|c|c|c|} \hline

Multiplet & Curve & Class & $g_{\Sigma}$ & $L_{\Sigma}$ & $L'_{\Sigma}$ \\
\hline \hline

$3\times {\bf 10}_{2,2}$ & $\Sigma_{M}^{1}$ & $4H+2E_2-E_1$ & 0 &
$\mathcal{O}_{\Sigma_{M}^{1}}(1)^{1/2}$ & $\mathcal{O}_{\Sigma_{M}^{1}}(1)$ \\
\hline

$3\times\bar{\bf 5}_{7,-1}$ & $\Sigma_{M}^{2}$ & $5H+3E_1-E_6$ & 0
&
$\mathcal{O}_{\Sigma_{M}^{2}}(-1)^{1/2}$ & $\mathcal{O}_{\Sigma_{M}^{2}}(1)^{5/14}$ \\
\hline

$3\times {\bf 1}_{7,5}$ & $\Sigma_{M}^{3}$ & $6H+3E_2-3E_3-2E_5$ &
0 &
$\mathcal{O}_{\Sigma_{M}^{3}}(1)^{1/2}$ & $\mathcal{O}_{\Sigma_{M}^{3}}(1)^{1/14}$ \\
\hline

$1\times {\bf 10}_{2,2}$ & $\Sigma_{H}^{1}$ & $2H-E_1-E_3-E_5$ & 0
&
$\mathcal{O}_{\Sigma_{H}^{1}}(1)^{1/6}$ & $\mathcal{O}_{\Sigma_{H}^{1}}(1)^{1/3}$ \\
\hline

$1\times\overline{\bf 10}_{-2,-2}$ & $\Sigma_{H}^{2}$ &
$2H-E_2-E_3-E_5$ & 0 &
$\mathcal{O}_{\Sigma_{H}^{2}}(-1)^{1/6}$ & $\mathcal{O}_{\Sigma_{H}^{2}}(-1)^{1/3}$ \\
\hline

$1\times {\bf 5}_{2,-4}$ & $\Sigma_{h}^{3}$ & $3H+E_1-E_3$ & 0 &
$\mathcal{O}_{\Sigma_{h}^{3}}(-1)^{1/6}$ & $\mathcal{O}_{\Sigma_{h}^{3}}(1)^{1/6}$ \\
\hline

$1\times \bar{\bf 5}_{7,-1}$ & $\Sigma_{h}^{4}$ & $H-E_2-E_3$ & 0
&
$\mathcal{O}_{\Sigma_{h}^{4}}(-1)^{1/6}$ & $\mathcal{O}_{\Sigma_{h}^{4}}(1)^{5/42}$ \\
\hline
\end{tabular}
\caption{An $SU(5)\times U(1)_X$ model from $G_{S}=SU(6)$, where
$L=\mathcal{O}_{S}(E_{1}-E_{2})^{1/6}$.} \label{SU6FSU5}
\end{center}
\end{table}

From the spectrum the matter fields $\bf 10$ and $\bar{\bf 5}$ are
from the curves that have different enhanced gauge groups, which
implies they are not unified in the same representation of a
higher rank gauge group, such as the $\bf 15$ of $SU(6)$.
Furthermore, we are not able to obtain the corresponding $U(1)_X$
charges of the matter after rotating the two charges of each
representation. These imply that a flipped $SU(5)$ gauge group is
not naturally embedded in $SU(6)$.  The approach to build an
$SU(5)\times U(1)_X$ from $G_S=SU(6)$ is not a success.


\subsubsection{$G_{S}=SO(10)$}

In this section we shall build the flipped $SU(5)$ model from the
bulk $G_{S}=SO(10)$. Again, we achieve this by extending the
spectrum of the $SU(5)$ model constructed from $G_S=SO(10)$ in
section (\ref{secsu5so10}). The $U(1)_S$ charges of the fields on
the curves should be conserved in the Yukawa couplings due to the
nontrivial bulk flux.  The Yukawa couplngs in the superpotential
are
\begin{eqnarray}
\mathcal{W}\supset {\bf 10}_{-3,-1} {\bf 10}_{-3,-1} {\bf
5}_{-2,2} +{\bf 10}_{-3,-1} \bar{\bf 5}_{-3,3} \bar{\bf 5}_{2,-2}
 +  \bar{\bf 5}_{-3,3}{\bf 5}_{-2,2} {\bf
1}_{-3,-5}+\cdots.
\end{eqnarray}
The matter singlet has 6 copies and is from the same curve
$\Sigma_{E_{6}}$ as the ${\bf 10}_M$.  The additional GUT Higgs
multiplets ${\bf 10}_{H}$ and $\overline{\bf 10}_{H}$ can be
engineered by the following calculation.

${\bf 10}_{H}$ has the same charge as the ${\bf 10}_{M}$ does, so
we also choose the enhanced gauge group of curve $\Sigma_{H}^{1}$
to be $G_{\Sigma_{H}^{1}}=E_{6}$. Let
$L_{\Sigma_{H}^{1}}=\mathcal{O}_{\Sigma_{H}^{1}}(f_{H}^{1})$ and
$L'_{\Sigma_{H}^{1}}=\mathcal{O}_{\Sigma_{H}^{1}}(g_{H}^{1})$. In
order to obtain the desired field content, it is required that
\[h^{0}(\Sigma_{H}^{1},K^{1/2}_{\Sigma_{H}^{1}}\otimes\mathcal{O}_{\Sigma_{H}^{1}}(-f_{H}^{1})
\otimes\mathcal{O}_{\Sigma_{H}^{1}}(-3g_{H}^{1}))=1,\]
\[h^{0}(\Sigma_{H}^{1},K^{1/2}_{\Sigma_{H}^{1}}\otimes\mathcal{O}_{\Sigma_{H}^{1}}(f_{H}^{1})
\otimes\mathcal{O}_{\Sigma_{H}^{1}}(3g_{H}^{1}))=0,\]
\[h^{0}(\Sigma_{H}^{1},K^{1/2}_{\Sigma_{H}^{1}}\otimes\mathcal{O}_{\Sigma_{H}^{1}}(3f_{H}^{1})
\otimes\mathcal{O}_{\Sigma_{H}^{1}}(-3g_{H}^{1}))=0,\]
\[h^{0}(\Sigma_{H}^{1},K^{1/2}_{\Sigma_{H}^{1}}\otimes\mathcal{O}_{\Sigma_{H}^{1}}(-3f_{H}^{1})
\otimes\mathcal{O}_{\Sigma_{H}^{1}}(3g_{H}^{1}))=0,\]
\[h^{0}(\Sigma_{H}^{1},K^{1/2}_{\Sigma_{H}^{1}}\otimes\mathcal{O}_{\Sigma_{H}^{1}}(-5f_{H}^{1})
\otimes\mathcal{O}_{\Sigma_{H}^{1}}(-3g_{H}^{1}))=0,\]
\[h^{0}(\Sigma_{H}^{1},K^{1/2}_{\Sigma_{H}^{1}}\otimes\mathcal{O}_{\Sigma_{H}^{1}}(5f_{H}^{1})
\otimes\mathcal{O}_{\Sigma_{H}^{1}}(3g_{H}^{1}))=0.\]

It is easy to see no solution satisfies all the conditions, which
means there exists exotic matter. We choose
$f_{H}^{1}=-\frac{1}{4}$ and $g_{H}^{1}=-\frac{1}{4}$ and the
field content is
\[1\times {\bf 10}_{-3,-1},\;\;\;2\times {\bf 1}_{-3,-5}.\]
Similarly, we take $\Sigma_{H}^{2}$ as a genus zero curve with
$G_{\Sigma_{H}^{2}}=E_{6}$ and let the line bundles be
$L_{\Sigma_{H}^{2}}=\mathcal{O}_{\Sigma_{H}^{2}}(f_{H}^{2})$ and
$L'_{\Sigma_{H}^{2}}=\mathcal{O}_{\Sigma_{H}^{2}}(g_{H}^{2})$.
Following the same process, we find that there is no solution for
all the conditions. So we set $f_{H}^{2}=\frac{1}{4}$ and
$g_{H}^{2}=\frac{1}{4}$ for a minimum content. The resulting field
content is
\[1\times \overline{\bf 10}_{3,1},\;\;\;2\times {\bf 1}_{3,5}\]
We summarize the result in Table \ref{SO(10)F5}.

\begin{table}[h]
\begin{center}
\renewcommand{\arraystretch}{1.25}
\begin{tabular}{|c|c|c|c|c|c|c|c|} \hline
Multiplet & Curve & Class & $g_{\Sigma}$ & $L_{\Sigma}$ & $L'_{\Sigma}$ \\
\hline \hline

$3\times {\bf 10}_{-3,-1}$ & $\Sigma_{M}^{1}$ & $4H+2E_1-E_2$ & 0
&
$\mathcal{O}_{\Sigma_{M}^{1}}(-1)^{3/4}$ & $\mathcal{O}_{\Sigma_{M}^{1}}(-1)^{3/4}$ \\
\hline

$3\times\bar{\bf 5}_{-3,3}$ & $\Sigma_{M}^{2}$ & $5H+3E_2-E_5$ & 0
&
$\mathcal{O}_{\Sigma_{M}^{2}}(1)^{3/4}$ & $\mathcal{O}_{\Sigma_{M}^{2}}(-1)^{1/4}$ \\
\hline

$3\times {\bf 1}_{-3,-5}$ & $\Sigma_{M}^{1}$ & $6H+3E_1-3E_4-2E_5$
& 0 &
$\mathcal{O}_{\Sigma_{M}^{1}}(-1)^{3/4}$ & $\mathcal{O}_{\Sigma_{M}^{1}}(-1)^{3/4}$ \\
\hline

$1\times {\bf 10}_{-3,-1}$ & $\Sigma_{H}^{1}$ & $2H-E_2-E_4-E_5$ &
0 &
$\mathcal{O}_{\Sigma_{H}^{1}}(-1)^{1/4}$ & $\mathcal{O}_{\Sigma_{H}^{1}}(-1)^{1/4}$ \\
\hline

$1\times\overline{\bf 10}_{3,1}$ & $\Sigma_{H}^{2}$ &
$2H-E_1-E_4-E_5$ & 0 &
$\mathcal{O}_{\Sigma_{H}^{2}}(1)^{1/4}$ & $\mathcal{O}_{\Sigma_{H}^{2}}(1)^{1/4}$ \\
\hline

$1\times {\bf 5}_{-2,2}$ & $\Sigma_{h}^{3}$ & $H-E_1-E_5$ & 0 &
$\mathcal{O}_{\Sigma_{h}^{3}}(1)^{1/4}$ & $\mathcal{O}_{\Sigma_{h}^{3}}(-1)^{1/4}$ \\
\hline

$1\times\bar{\bf 5}_{2,-2}$ & $\Sigma_{h}^{4}$ & $H-E_2-E_5$ & 0 &
$\mathcal{O}_{\Sigma_{h}^{4}}(-1)^{1/4}$ & $\mathcal{O}_{\Sigma_{h}^{4}}(1)^{1/4}$ \\
\hline

\end{tabular}
\caption{An $SU(5)\times U(1)_X$ model from $G_{S}=SO(10)$, where
$L=\mathcal{O}_{S}(E_{1}-E_{2})^{1/4}$}\label{SO(10)F5}
\end{center}
\end{table}

The $U(1)_S$ charges in the spectrum are consistent with the
$U(1)_X$ charges, which is natural since $SU(5)\times U(1)_X$ is
embedded in $SO(10)$. However, to make $U(1)_X$ massless we have
to rotate the $U(1)$ gauge groups to satisfy the constraints from
the Green-Schwarz mechanism in a global picture. In addition, we
are not able to avoid a few copies of exotic singlets. This model
includes all the terms of the generic superpotential $\mathcal{W}$
of $SU(5)\times U(1)_X$ stated in Eq.~(\ref{superFSU5}).

From the first case, we find the generic structure of $G_S=SU(6)$
cannot produce a flipped $SU(5)$ model due to the inconsistent
charges of the fermion and Higgs fields.  It is difficult to
construct a flipped $SU(5)$ model unless we are able to turn on a
line bundle to break $G_{\Sigma}$ to an $SO(10)$ gauge group.

In the second case, the $SU(5)\times U(1)_X$ model from $SO(10)$
is similar to the constructions in \cite{Beasley:2008kw} and
\cite{Jiang:2008yf}.  In our model, the curves in the spectrum
have alternate classes. The nontrivial bulk fluxes on the curves
are turned on so we can study the substructure of $\bf 16$ from
$SO(10)$. $\bar{\bf 5}$ and $\bf 10$ are not on the same curve,
while $\bf 10$ still forms a $\bf 10\,10\,5$ coupling but
$\bar{\bf 5}$ gets rid of the coupling ${\bf \bar 5\, \bar 5\,
5}_h$. The $U(1)_S$ charges are consistent with the $U(1)_X$
charges. This implies the bulk $SO(10)$ is corresponding to the
$SO(10)$ GUT which is the higher unification of the flipped
$SU(5)$.

Again the self-intersecting geometry can be introduced to obtain a
rank three Yukawa mass structure, and we can also construct a
flipped $SU(5)$ model by splitting chiral fermions on two
different matter curves \cite{Font:2008id}.


\subsection{$SO(10)$ GUT}

In this section we shall discuss the $SO(10)$ GUT from the
breaking of a higher rank bulk gauge group.  There are two
possible choices, $G_{S}=SO(12)$ and $G_{S}=E_6$.

\subsubsection{$G_{S}=SO(12)$}
Consider seven-branes wrapping on $S$ where $G_{S}=SO(12)$. There
exist the following breaking patterns from the enhanced adjoints
of the curves:
\begin{equation}
\begin{array}{c@{}c@{}l@{}c@{}l}
SO(14) &~\rightarrow~& SO(12)_S\times U(1) &~\rightarrow~& SO(10)\times U(1)\times U(1)_{S}\\
 91 &~\rightarrow~& 66_0+1_0&~\rightarrow~& 45_{0,0}+1_{0,0}+10_{0,2}+
 \overline{10}_{0,-2}+1_{0,0}\\
 & &+12_2+\overline{12}_{-2} & &+(10_{2,0}+1_{2,2}+1_{2,-2}+c.c.)
\end{array}
\end{equation}
\begin{equation}
\begin{array}{c@{}c@{}l@{}c@{}l}
E_7 &~\rightarrow~ & SO(12)_S\times U(1) &~\rightarrow~& SO(10)\times U(1)\times U(1)_S\\
 133 &~\rightarrow~ & 66_0+1_0+{\it 1_{\pm2}}&~\rightarrow~&
45_{0,0}+2\times 1_{0,0}+{10_{0,2}}+ \overline{10}_{0,-2}+{\it1_{\pm2,0}}\\
 & & +32'_1+{\it {32'}_{-1}} & &+16_{1,-1}+\overline{16}_{1,1}+\it 16_{-1,-1}+\overline{16}_{-1,1}
\end{array}
\end{equation}

To obtain a $\bf 16\,16\,10$ coupling, the $\bf 10$ can only be
from the bulk due to the conservation of the $U(1)_S$ charges, and
it implies that the coupling is a $\Sigma\Sigma S$-type instead of
a $\Sigma\Sigma\Sigma$-type. From the above breaking patterns, the
possible choices are ${\bf 16}_{1,-1} {\bf 16}_{1,-1} {\bf
10}_{0,2}$, ${\bf 16}_{-1,-1} {\bf 16}_{-1,-1} {\bf 10}_{0,2}$,
and ${\bf 16}_{1,-1} {\bf 16}_{-1,-1} {\bf 10}_{0,2}$, and we take
the first as an example whose superpotential is:
\begin{equation}
\mathcal{W}\supset  {\bf 16}_{1,-1} {\bf 16}_{1,-1} {\bf
10}_{0,2}+\cdots
\end{equation}
The corresponding Yukawa coupling pattern on the double enhanced
point can be found in Eq. (\ref{so12161610}).

We choose a genus zero curve $\Sigma_{M}^{1}$ with
$G_{\Sigma_{M}^{1}}=E_{7}$ and let
$L_{\Sigma_{M}^{1}}=\mathcal{O}_{\Sigma_{M}^{1}}(h_{M}^{1})$ and
$L'_{\Sigma_{M}^{1}}=\mathcal{O}_{\Sigma_{M}^{1}}(k_{M}^{1})$. In
order to get the desired field content, it is required that
\[h^{0}(\Sigma_{M}^{1},K^{1/2}_{\Sigma_{M}^{1}}\otimes\mathcal{O}_{\Sigma_{M}^{1}}(-h_{M}^{1})\mathcal{O}_{\Sigma_{M}^{1}}(k_{M}^{1}))=3,\]
\[h^{0}(\Sigma_{M}^{1},K^{1/2}_{\Sigma_{M}^{1}}\otimes\mathcal{O}_{\Sigma_{M}^{1}}(h_{M}^{1})\mathcal{O}_{\Sigma_{M}^{1}}(-k_{M}^{1}))=0,\]
\[h^{0}(\Sigma_{M}^{1},K^{1/2}_{\Sigma_{M}^{1}}\otimes\mathcal{O}_{\Sigma_{M}^{1}}(-h_{M}^{1})
\otimes\mathcal{O}_{\Sigma_{M}^{1}}(-k_{M}^{1}))=0,\]
\[h^{0}(\Sigma_{M}^{1},K^{1/2}_{\Sigma_{M}^{1}}\otimes\mathcal{O}_{\Sigma_{M}^{1}}(h_{M}^{1})
\otimes\mathcal{O}_{\Sigma_{M}^{1}}(k_{M}^{1}))=0.\] The unique
solution is $h_{M}^{1}=-\frac{3}{2}$ and $k_{M}^{1}=\frac{3}{2}$,
so the resulting field content is
\[3\times {\bf 16}_{1,-1}.\]
The Higgs multiplet ${\bf 10}_{0,2}$ is from the bulk. By Eq.
(\ref{EulerChar}), we obtain
\[N_{{\bf 10}_{2}}=1,\;N_{\overline{{\bf 10}}_{-2}}=0\]
where $L=\mathcal{O}_{S}(E_{1}-E_{2}-E_{3})^{1/2}$ has been used.
Note that in this case, we change the polarization to be
$J_S=AH-2E_1-\sum^{8}_{i=2}E_i$ so that BPS equation (\ref{BPS})
still holds. The spectrum is shown in Table \ref{SO(12)SO(10)}.

\begin{table}[h]
\begin{center}
\renewcommand{\arraystretch}{1}
\begin{tabular}{|c|c|c|c|c|c|c|} \hline

Multiplet & Curve & Class & $g_{\Sigma}$ & $L_{\Sigma}$ & $L'_{\Sigma}$ \\
\hline \hline

$3\times {\bf 16}_{1,-1}$ & $\Sigma_{M}^{1}$ & $3H+E_1-E_2-E_3$ &
0 & $\mathcal{O}_{\Sigma_{M}^{1}}(-1)^{3/2}$ & $\mathcal{O}_{\Sigma_{M}^{1}}(1)^{3/2}$ \\
\hline

\end{tabular}
\caption{An $SO(10)$ GUT model from $G_S=SO(12)$, where
$L=\mathcal{O}_{S}(E_{1}-E_{2}-E_{3})^{1/2}$ and Higgs $\bf 10$ is
from the bulk.}\label{SO(12)SO(10)}
\end{center}
\end{table}


\subsubsection{$G_{S}=E_{6}$}

In the case of $G_S=E_{6}$, $E_{6}$ is broken into $SO(10)\times
U(1)_S$ by nontrivial fluxes on the bulk. In order to avoid chiral
matter on the bulk, we choose a supersymmetric line bundle
$L=\mathcal{O}_{S}(E_{1}-E_{2})^{1/3}$ over $S$. By doing so, all
chiral matter on the bulk disappears, $i.e.$ $N_{{\bf
16}_{0,-3}}=N_{{\bf\overline{16}}_{0,3}}=0$, which means that all
the chiral fields are localized on the curves. The possible
breaking chain and the matter content from the enhanced adjoint of
the curve is
\begin{equation}
\begin{array}{c@{}c@{}l@{}c@{}l}
E_7&\rightarrow&E_6\times U(1) &\rightarrow&
SO(10)\times U(1)\times U(1)_S \\
133 &\rightarrow & ~78_0+1_0 & \rightarrow &
~45_{0,0}+1_{0,0}+1_{0,0} +16_{0,-3}+\overline{16}_{0,3} \\
& & +27_2+\overline{27}_{-2} & & +(16_{2,1}+10_{2,-2}+1_{2,4}
+c.c.) \\
\end{array}
\end{equation}
From the breaking pattern we find the Yukawa coupling in the
superpotential is $\Sigma\Sigma \Sigma$-type instead of
$\Sigma\Sigma S$-type:
\begin{equation}
\mathcal{W}\supset  {\bf 16}_{2,1} {\bf 16}_{2,1} {\bf
10}_{2,-2}+\cdots.
\end{equation}
The corresponding Yukawa coupling pattern on the double enhanced
point can be found in Eq. (\ref{e6161610}).

Consider $\Sigma_{M}^{1}$ a pinched curve of genus zero with
$G_{\Sigma_{M}^{1}}=E_{7}$ and let
$L_{\Sigma_{M}^{1}}=\mathcal{O}_{\Sigma_{M}^{1}}({\tilde
h}_{M}^{1})$ and
$L'_{\Sigma_{M}^{1}}=\mathcal{O}_{\Sigma_{M}^{1}}({\tilde
k}_{M}^{1})$. In order to get the desired field content, it is
required that
\[h^{0}(\Sigma_{M}^{1},K^{1/2}_{\Sigma_{M}^{1}}\otimes\mathcal{O}_{\Sigma_{M}^{1}}({\tilde
h}_{M}^{1})\otimes \mathcal{O}_{\Sigma_{M}^{1}}(2{\tilde
k}_{M}^{1}))=3,\]
\[h^{0}(\Sigma_{M}^{1},K^{1/2}_{\Sigma_{M}^{1}}\otimes\mathcal{O}_{\Sigma_{M}^{1}}(-{\tilde
h}_{M}^{1})\otimes \mathcal{O}_{\Sigma_{M}^{1}}(-2{\tilde
k}_{M}^{1}))=0,\]
\[h^{0}(\Sigma_{M}^{1},K^{1/2}_{\Sigma_{M}^{1}}\otimes\mathcal{O}_{\Sigma_{M}^{1}}(-2{\tilde
h}_{M}^{1}) \otimes\mathcal{O}_{\Sigma_{M}^{1}}(2{\tilde
k}_{M}^{1}))=0,\]
\[h^{0}(\Sigma_{M}^{1},K^{1/2}_{\Sigma_{M}^{1}}\otimes\mathcal{O}_{\Sigma_{M}^{1}}(2{\tilde
h}_{M}^{1}) \otimes\mathcal{O}_{\Sigma_{M}^{1}}(2{\tilde
k}_{M}^{1}))=0,\]
\[h^{0}(\Sigma_{M}^{1},K^{1/2}_{\Sigma_{M}^{1}}\otimes\mathcal{O}_{\Sigma_{M}^{1}}(4{\tilde
h}_{M}^{1}) \otimes\mathcal{O}_{\Sigma_{M}^{1}}(2{\tilde
k}_{M}^{1}))=0,\]
\[h^{0}(\Sigma_{M}^{1},K^{1/2}_{\Sigma_{M}^{1}}\otimes\mathcal{O}_{\Sigma_{M}^{1}}(-4{\tilde
h}_{M}^{1}) \otimes\mathcal{O}_{\Sigma_{M}^{1}}(-2{\tilde
k}_{M}^{1}))=0.\] Since there is no solution for all the
conditions, it implies there exists exotic matter. We choose
${\tilde h}_{M}^{1}=1$ and ${\tilde k}_{M}^{1}=1$, so the
resulting field content is
\[3\times {\bf 16}_{2,1},\;\;\;6\times {\bf 1}_{2,4}.\]
We choose $\Sigma_{H}^{1}$ to be a genus zero curve with
$G_{\Sigma_{H}^{1}}=E_{7}$. Let the line bundles on
$\Sigma_{H}^{1}$ be
$L_{\Sigma_{H}^{1}}=\mathcal{O}_{\Sigma_{H}^{1}}({\tilde
h}_{H}^{1})$ and
$L'_{\Sigma_{H}^{1}}=\mathcal{O}_{\Sigma_{H}^{1}}({\tilde
k}_{H}^{1})$. Again, there is no solution for all the conditions.
We then choose ${\tilde h}_{H}^{1}=-\frac{1}{3}$ and ${\tilde
k}_{H}^{1}=\frac{1}{6}$, so the resulting field content is
\[1\times {\bf 10}_{2,-2},\;\;\;1\times {\bf 1}_{-2,-4}.\]
We summarize the result in Table \ref{E_6SO(10)}.

\begin{table}[h]
\begin{center}
\renewcommand{\arraystretch}{1}
\begin{tabular}{|c|c|c|c|c|c|c|} \hline
Multiplet & Curve & Class & $g_{\Sigma}$ & $L_{\Sigma}$ & $L'_{\Sigma}$ \\
\hline \hline

$3\times {\bf 16}_{2,1}$ & $\Sigma_{M}^{1}$ & $4H+2E_2-E_1$ & 0 &
$\mathcal{O}_{\Sigma_{M}^{1}}(1)$ & $\mathcal{O}_{\Sigma_{M}^{1}}(1)$ \\
\hline

$1\times {\bf 10}_{2,-2}$ & $\Sigma_{H}^{1}$ & $H-E_2-E_3$ & 0 &
$\mathcal{O}_{\Sigma_{H}^{2}}(-1)^{1/3}$ & $\mathcal{O}_{\Sigma_{H}^{1}}(1)^{1/6}$ \\
\hline

\end{tabular}
\caption{An $SO(10)$ GUT model from $G_S=E_6$, where
$L=\mathcal{O}_{S}(E_{1}-E_{2})^{1/3}$}\label{E_6SO(10)}
\end{center}
\end{table}

In these models, the fluxes are nontrivial on all the curves in
order to break the gauge group into $SO(10)$. In the first
example, the fields come from both the bulk and the curve, while
in the second the fields are from the curves.

To solve the doublet-triblet problem, we may consider the
Dimopoulos-Wilczek mechanism \cite{DimoWil}.  There are several
choices of Higgs fields to break the $SO(10)$ gauge group, but
they are absent in these models. For example, we do not have {\bf
210}, {\bf 210}, and ${\bf 126}+\overline{\bf 126}$ to break the
gauge group to the $SU(5)$ GUT or MSSM-like model
\cite{Fukuyama:2004ps}. However, the configurations of the
non-Abelian instanton broken into a product of $U(1)$s may take
the work \cite{Beasley:2008kw}. The possible breaking pattern is
\begin{eqnarray}
&SO(10)\times U(1)_{S_a} \rightarrow & SU(5)\times
U(1)_{S_b}\times U(1)_{S_a} \rightarrow
SU(3)\times SU(2)\times U(1)^3_S, \nonumber \\
{\rm or~}&SO(10)\times U(1)_{S_a} \rightarrow & SU(2)\times
SU(2)\times SU(4)\times U(1)_{S_a} \nonumber \\ && \rightarrow
SU(2)\times SU(2)\times SU(3)\times U(1)^2_S.
\end{eqnarray}

\section{Conclusion}

In this paper we construct examples of $SU(5)$, $SU(5)\times
U(1)_X$, and $SO(10)$ GUT local models from $G_S$ which is one
rank higher than these GUT gauge groups in the F-theory
configuration. The bulk flux is nontrivial on all the curves to
break $G_S$ down to the GUT gauge group.  We can study the the
unification of the GUT gauge groups to higher rank gauge groups in
string theory. There is no GUT adjoint representation on a del
Pezzo surface, but it is still possible to break the GUT gauge
groups to the SM gauge group by introducing Abelian instanton
configurations on the bulk \cite{Beasley:2008kw}.

We demonstrate how to obtain a model of $SU(5)$ Georgi-Glashow
from $G_S=SU(6)$. In this model we are able to obtain three copies
of quarks and leptons in the $\bf 10$ and $\bar{\bf 5}$
representations and one copy of the Higgs fields ${\bf 5}_H$ and
$\bar{\bf 5}_H$. Due to the $U(1)_S$ charge structure when
breaking $SU(6)$ to $SU(5)$, the up-type Higgs and down-type Higgs
are not charge conjugates. To obtain the $\mu$ term a mixture
state for the up-type Higgs from two curves may be considered and
further studied.  In these models $SU(5)$ descends from an $SU(6)$
unification. In the example of $SU(5)$ from $G_S=SO(10)$, the
$U(1)_S$ charges are consistent in each term of superpotential,
and we can see it is natural to embed $SU(5)$ into $SO(10)$. In
our examples the matter $\bf 10$ is either from a curve or two
independent curves from which it is possible to use the left-right
mechanism to generate rank three mass matrices elegantly as shown
in \cite{Font:2008id}. In these $SU(5)$ models we can avoid rapid
proton decay by separating the up and down type Higgs from
vector-like pairs, and the generic doublet-triplet splitting
problem may be controlled when GUT breaks down to MSSM by the
additional $U(1)$ from the instanton.

We also try to construct a flipped $SU(5)$ model from $G_S=SU(6)$
and $G_S=SO(10)$.  However we are not able to find a consistent
set of $U(1)_X$ charges for the matter content in the model with
$G_S=SU(6)$.  This implies it is not natural to embed an
$SU(5)\times U(1)_X$ GUT into an $SU(6)$ gauge group.  In the
example of $G_S=SO(10)$ the fermion spectrum is similar to what we
obtained in the case of $SU(5)$ Georgi-Glashow, with an additional
pair of ${\bf 10}_H$ and ${\overline{\bf 10}}_H$ Higgs fields. The
$U(1)_S$ charges are consistent with the $U(1)_X$ charges which
implies $SO(10)$ is a more natural unification from $SU(5)\times
U(1)_X$.  For a massless $U(1)_X$ one may have to refer to the
global picture. The model construction is similar to that studied
in \cite{Jiang:2008yf}, but $\bar{\bf 5}$ fermion is from a
different curve from ${\bf 10}$. One advantage of the model is
that we can avoid the $\bar{\bf 5}\, \bar{\bf 5}\, \bf 5$ coupling
in the superpotential.

In addition, we demonstrate how to obtain models of an $SO(10)$
GUT from $G_S=SO(12)$ and $G_S=E_6$. In the case of $G_S=SO(12)$,
the ${\bf 10}_H$ field is from the bulk so the matter Yukawa
coupling is a $\Sigma\Sigma S$ type, while in the $G_S=E_6$ case,
all the matter fields are from bi-fundamental representations.
There is no $SO(10)$ adjoint $\Phi_{\bf 45}$ for a coupling such
as $\Phi_{\bf 45}\, {\bf 16}_H\, \overline{\bf 16}_H $, however
one may consider introducing the instanton configuration to break
the GUT gauge symmetry.

The singularity types on the fibers are corresponding to the gauge
groups on the seven-branes in F-theory.  The introduction of
fluxes can be regarded as resolutions of the singularities, and
then we are able to analyze the fluxes via Cartan subalgebra
\cite{Katz:1996xe}. There then arises an interesting question that
whether the enhanced gauge group on the curve breaks to a gauge
group different from the original bulk gauge group when the line
bundle is turned on. It may result in interesting gauge group
configurations on the curves.

F-theory has captured attention recently for its non-perturbative
configuration and elegant way of constructing the matter spectrum
of a local model. The next step is probably to find out the global
constraints for building realistic models. Other topics, like
supersymmtry breaking, non-abelian gauge fluxes for gauge group
breaking to MSSM, and explicit examples of del Pezzo surfaces for
GUT models are interesting and worthy of study in the future.

\renewcommand{\thesection}{}
\section{\hspace{-1cm} Acknowledgments}

We would like to thank K. Becker, C. Bertinato, B. Dutta, J.
Heckman, and J. Marsano for valuable communications and
discussions. The work of CMC is supported in part by the
Mitchell-Heep Chair in High Energy Physics. The work of YCC is
supported in part by the NSF grant PHY-0555575 and Texas A\&M
University.

\newpage

\appendix{\bf\Large \hspace{-.85cm} Appendix}

\renewcommand{\theequation}{\thesection.\arabic{equation}}
\setcounter{equation}{0}

\section{Del Pezzo Surfaces} In this section we shall briefly
review the geometric properties of del Pezzo surfaces. Del Pezzo
surface $dP_{k},\;k\leq 8$ is defined by blowing up $k$ generic
points of $\mathbb{P}^2$ or $\mathbb{P}^{1}\times\mathbb{P}^{1}$.
The divisors on $dP_{k}$ can be generated by $H$ and $E_{i}$,
where $H$ is a hyperplane divisor, and $E_{i}$ is an exceptional
divisor from blowing-up and is isomorphic to $\mathbb{P}^{1}$. The
intersecting numbers are
\[H\cdot H=1,\;\;\:E_{i}\cdot E_{j}=-\delta_{ij},\;\;\;H\cdot E_{i}=0.\]
The canonical divisor on $dP_{k}$ is given by
\begin{equation}
K_{dP_{k}}=-c_{1}(dP_{k})=-3H+\sum_{i=1}^{k}E_{i}.
\end{equation}
The genus of the curve $\mathcal{C}$ within $dP_{k}$ can be
calculated by the the formula
\[\mathcal{C}\cdot (K_{dP_{k}}+\mathcal{C})=2g-2.\]
For a large volume limit, given a line bundle $L$ on $dP_{k}$ and
\begin{equation}
c_{1}(L)=\sum_{i=1}^{k}a_{i}E_{i},
\end{equation}
where $a_{i}a_{j}<0$ for some $i\neq j $, there exits a parametric
family of K\"ahler classes $J_{dP_{k}}$ over $dP_{k}$ constructed
as \cite{Beasley:2008dc}
\begin{equation}
J_{dP_{k}}=AH-\sum_{i=1}^{k}b_{i}E_{i},
\end{equation}
where $\sum_{k}a_{k}b_{k}=0$ and $A\gg b_{i}>0$. By the
construction, it is easy to see that the line bundle $L$ solves
the BPS equation $J_{dP_k}\wedge c_{1}(L)=0$.

\section{Resolutions of the Triplet Intersections}
\subsection{$SU(5)$ GUT Model}
\setcounter{equation}{0}

For $SU(5)$ GUT model, we consider $G_{S}$ and $G_{p}$ to be of
rank five and seven, respectively. In general, we have
$G_{p}=SU(8)$, $SO(14)$ or $E_{7}$. Here we only consider the
group theory decompositions of $ADE$ type. It is straightforward
to get the following resolutions \cite{Slansky:1981yr}:

\paragraph{$G_{p}=SU(8):$}
\begin{equation}
\small
\begin{array}{c@{}c@{}l@{}c@{}l@{}c@{}l}
SU(8)&\rightarrow&SU(7)\times U(1) &\rightarrow&
SU(6)_S\times U(1)^2 &\rightarrow& SU(5)\times U(1)^2\times U(1)_S\\
63 &\rightarrow & ~48_0+1_0 & \rightarrow &
~35_{0,0}+1_{0,0}+1_{0,0} &\rightarrow &
~24_{0,0,0}+3\times 1_{0,0,0}+5_{0,0,6}+\bar 5_{0,0,-6}\\
& & & & +6_{0,-7}+\bar 6_{0,7} & & +(5_{0,-7,1}+1_{0,-7,-5}+c.c.) \\
& & +7_8+\bar 7_{-8} & & +(6_{8,-1}+1_{8,6}+c.c.)&
&+(5_{8,-1,1}+1_{8,-1,-5}+1_{8,6,0}+c.c.)
\end{array}
\end{equation}
\paragraph{$G_{p}=SO(14):$}
\begin{equation}
\small
\begin{array}{c@{}c@{}l@{}c@{}l@{}c@{}l}
SO(14)&\rightarrow&SO(12)\times U(1) &\rightarrow&
SO(10)\times U(1)^2 &\rightarrow& SU(5)\times U(1)^2\times U(1)_S\\
91 &\rightarrow & ~66_0+1_0 & \rightarrow &
~45_{0,0}+1_{0,0}+1_{0,0} &\rightarrow & ~24_{0,0,0}+3\times
1_{0,0,0}+{10_{0,0,4}} +\overline{10}_{0,0,-4}\\ & & & &
+10_{0,2}+\overline{10}_{0,-2} & &
+(5_{0,2,2}+{\bar 5_{0,2,-2}}+c.c.) \\
& & +12_2+\overline{12}_{-2} & & +(10_{2,0}+1_{2,2}+1_{2,-2}&
&+(5_{2,0,2}+\bar 5_{2,0,-2}+1_{2,2,0}+1_{2,-2,0} \\ & & & &
+~c.c.) & & +~c.c.)
\end{array} \label{so10105-5-}
\end{equation}
\begin{equation}
\small
\begin{array}{c@{}c@{}l@{}c@{}l@{}c@{}l}
SO(14)&\rightarrow&SO(12)\times U(1) &\rightarrow&
SU(6)\times U(1)^2 &\rightarrow& SU(5)\times U(1)^2\times U(1)_S\\
91 &\rightarrow & ~66_0+1_0 & \rightarrow &
~35_{0,0}+1_{0,0}+1_{0,0}
&\rightarrow & ~24_{0,0,0}+3\times 1_{0,0,0}+5_{0,0,6}+{\bar 5_{0,0,-6}}\\
& & & & +15_{0,2}+\overline{15}_{0,-2} & &
+({10_{0,2,2}}+5_{0,2,-4}+c.c.) \\
& & +12_2+\overline{12}_{-2} & & +(6_{2,1}+\bar 6_{2,-1}+c.c.) &
&+(5_{2,1,1}+1_{2,1,-5}+{\bar 5_{2,-1,-1}}+1_{2,-1,5}+c.c.)
\end{array}
\end{equation}
\begin{equation}
\small
\begin{array}{c@{}c@{}l@{}c@{}l@{}c@{}l}
SO(14)&\rightarrow&SU(7)\times U(1) &\rightarrow&
SU(6)\times U(1)^2 &\rightarrow& SU(5)\times U(1)^2\times U(1)_S\\
91 &\rightarrow & ~48_0+1_0 & \rightarrow &
~35_{0,0}+1_{0,0}+1_{0,0}
&\rightarrow & ~24_{0,0,0}+3\times 1_{0,0,0}+5_{0,0,6}+\bar 5_{0,0,-6}\\
& & & & +6_{0,-7}+\bar 6_{0,7} & & +(5_{0,-7,1}+1_{0,-7,-5}+c.c.) \\
& & +21_4+\overline{21}_{-4} & & +(15_{4,-2}+6_{4,5}+c.c.)& &+(
10_{4,-2,2} +5_{4,-2,-4}+5_{4,5,1}+1_{4,5,-5}+c.c.)
\end{array} \label{su6105-5-}
\end{equation}
\paragraph{$G_{p}=E_{7}:$}
\begin{equation}
\small
\begin{array}{c@{}c@{}l@{}c@{}l@{}c@{}l}
E_7&\rightarrow&E_6\times U(1) &\rightarrow&
SO(10)\times U(1)^2 &\rightarrow& SU(5)\times U(1)^2\times U(1)_S\\
133 &\rightarrow & ~78_0+1_0 & \rightarrow & ~45_{0,0}+2\times
1_{0,0} &\rightarrow &
~24_{0,0,0}+3\times 1_{0,0,0}+{ 10_{0,0,4}}+\overline{10}_{0,0,-4}\\
& & & & +16_{0,-3}+\overline{16}_{0,3} & &
+(10_{0,-3,-1}+\bar 5_{0,-3,3}+1_{0,-3,-5}+c.c.) \\
& & +27_2+\overline{27}_{-2} & & +(16_{2,1}+10_{2,-2}+1_{2,4}&
&+(10_{2,1,-1}+\bar 5_{2,1,3}+1_{2,1,-5} \\
& & & & +~c.c.) & & +~5_{2,-2,2}+\bar 5_{2,-2,-2}+1_{2,4,0}+c.c.)
\end{array} \label{so1010105}
\end{equation}
\begin{equation}
\small
\begin{array}{c@{}c@{}l@{}c@{}l@{}c@{}l}
E_7&\rightarrow&E_6\times U(1) &\rightarrow&
SU(6)\times U(1)^2 &\rightarrow& SU(5)\times U(1)^2\times U(1)_S\\
133 &\rightarrow & ~78_0+1_0 & \rightarrow & ~35_{0,0}+2\times
1_{00}+ {\it 1_{0,\pm2}} &\rightarrow &
~24_{0,0,0}+3\times 1_{0,0,0}+ {\it 1_{0,\pm2,0}}+5_{0,0,6}+{\bar 5_{0,0,-6}}\\
& & & & +20_{0,1}+{\it {20}_{0,-1}} & &
+10_{0,1,-3}+\overline{10}_{0,1,3}+{\it 10_{0,-1,-3}+\overline{10}_{0,-1,3}} \\
& & +27_2+\overline{27}_{-2} & & +(15_{2,0}+{\bar 6}_{2,1}+{\it
{\bar 6}_{2,-1}}&
&+({10_{2,0,2}}+ 5_{2,0,-4}+{\bar 5}_{2,1,-1}+1_{2,1,5} \\
& & & & +~c.c.) & & +~{\it {\bar 5}_{2,-1,-1}+1_{2,-1,5}}+c.c.)
\end{array}
\end{equation}
\begin{equation}
\small
\begin{array}{c@{}c@{}l@{}c@{}l@{}c@{}l}
E_7&\rightarrow&SO(12)\times U(1) &\rightarrow&
SO(10)\times U(1)^2 &\rightarrow& SU(5)\times U(1)^2\times U(1)_S\\
133 &\rightarrow & ~66_0+1_0+{\it 1_{\pm2}} & \rightarrow &
~45_{0,0}+2\times 1_{0,0}+ {\it 1_{\pm2,0}} &\rightarrow &
~24_{0,0,0}+3\times 1_{0,0,0}+ {\it 1_{\pm2,0,0}}+({10_{0,0,4}}+c.c) \\
& & & & +10_{0,2}+ \overline{10}_{0,-2} & &
+(5_{0,2,2}+\bar{5}_{0,2,-2}+c.c.) \\
& & +32'_1 & & +16_{1,-1}+\overline{16}_{1,1}&
&+{10_{1,-1,-1}}+\bar 5_{1,-1,3}+1_{1,-1,-5} \\
& & & & & &
+{\overline{10}_{1,1,1}}+ 5_{1,1,-3}+1_{1,1,5} \\
& & +{\it {32'}_{-1}} & & +\it 16_{-1,-1}+\overline{16}_{-1,1} & &
+\it {10_{-1,-1,-1}}+\bar 5_{-1,-1,3}+1_{-1,-1,-5} \\
& & & & & &
+\it {\overline{10}_{-1,1,1}}+ 5_{-1,1,-3}+1_{-1,1,5} \\
\end{array}
\end{equation}
\begin{equation}
\small
\begin{array}{c@{}c@{}l@{}c@{}l@{}c@{}l}
E_7&\rightarrow&SO(12)\times U(1) &\rightarrow&
SU(6)\times U(1)^2 &\rightarrow& SU(5)\times U(1)^2\times U(1)_S\\
133 &\rightarrow & ~66_0+1_0+{\it 1_{\pm2}} & \rightarrow &
~35_{0,0}+2\times 1_{00}+ {\it 1_{\pm2,0}} &\rightarrow &
~24_{0,0,0}+3\times 1_{0,0,0} + {\it 1_{\pm2,0,0}}+({5_{0,0,6}}+c.c.)\\
& & & & +15_{0,2}+ \overline{15}_{0,-2} & &
+(10_{0,2,2}+5_{0,2,-4}+c.c.) \\
& & +32'_1 & & +15_{1,-1}+\overline{15}_{1,1}+1_{1,\pm3}&
&+{10_{1,-1,2}}+ 5_{1,-1,-4}+\overline{10}_{1,1,-2}+\bar{5}_{1,1,4}+1_{1,\pm3,0} \\
& & +{\it {32'}_{-1}}& & {\it
+15_{-1,-1}+\overline{15}_{-1,1}+1_{-1,\pm3}} & & +\it
{10_{-1,-1,2}}+
5_{-1,-1,-4}+\overline{10}_{-1,1,-2}+\bar{5}_{-1,1,4}\\
&&&&&&+\it 1_{-1,\pm3,0}
\end{array}   \label{su610105}
\end{equation}

\subsubsection{$G_{S}=SU(6)$}

For $G_{S}=SU(6)$, we have the following enhancement patterns
\[SU(6)\rightarrow SU(7)\rightarrow SU(8)\]
with $G_{p}=SU(8)$,
\[SU(6)\rightarrow SO(12)\rightarrow SO(14)\]
with $G_{p}=SO(14)$,
\[SU(6)\rightarrow E_{6}\rightarrow E_{7}\]
with $G_{p}=E_{7}$, and
\[SU(6)\rightarrow SO(12)\rightarrow E_{7}\]
with $G_{p}=E_{7}$.

In this case, we only get the coupling $\bf 5\,\bar{5} \,1$ at
$G_{p}=SU(8)$, and from $G_{p}=SO(14)$ we are able to obtain
couplings $\bf 10\, \bar{5}\,\bar{5}$ and $\bf 5\,\bar{5}\, 1$. In
addition, we also get the most important one, $\bf 10\,10\,5$,
from $G_{p}=E_{7}$.

\subsubsection{$G_{S}=SO(10)$}

For $G_{S}=SO(10)$, we have following enhancement patterns
\[SO(10)\rightarrow SO(12)\rightarrow SO(14)\]
with $G_{p}=SO(14)$,
\[SO(10)\rightarrow E_{6}\rightarrow E_{7}\]
with $G_{p}=E_{7}$, and
\[SO(10)\rightarrow SO(12)\rightarrow E_{7}\]
with $G_{p}=E_{7}$

In this case, we have the couplings $\bf 10\, \bar{5}\,\bar{5}$
and $\bf 5\, \bar{5}\,1$ from $G_{p}=SO(14)$, and we can also
obtain the most important one, $\bf 10\,10\,5$, from
$G_{p}=E_{7}$. Note that we are not able to get $G_{p}=SU(8)$
which gives rise to the coupling $\bf 5\,\bar{5}\,1$. Fortunately,
this coupling can found in $G_{p}=SO(14)$ or $G_{p}=E_{7}$
instead.

\subsection{$SO(10)$ GUT Model}

For $SO(10)$ GUT model, we consider $G_{S}$ and $G_{p}$ to be of
rank six and eight, respectively.  Here we only consider the case
of $G_p=SO(16)$ and $E_{8}$. It is straightforward to get the
following resolutions:

\paragraph{$G_{p}=SO(16):$}

\begin{equation}
\small
\begin{array}{c@{}c@{}l@{}c@{}l@{}c@{}l}
SO(16)&\rightarrow&SO(14)\times U(1) &\rightarrow&
SO(12)\times U(1)^2 &\rightarrow& SO(10)\times U(1)^2\times U(1)_S\\
120 &\rightarrow & ~91_0+1_0 & \rightarrow &
~66_{0,0}+1_{0,0}+1_{0,0}
&\rightarrow & ~45_{0,0,0}+3\times 1_{0,0,0}+10_{0,0,2}+{\overline{10}_{0,0,-2}} \\
& & & & +12_{0,2}+\overline{12}_{0,-2} & & +(10_{0,2,0}+1_{0,2,2}+1_{0,2,-2}+c.c.) \\
& & +14_2+\overline{14}_{-2} & & +(12_{2,0}+1_{2,2}+1_{2,-2}&
&+(10_{2,0,0}+1_{2,0,2}+1_{2,0,-2}  \\ & & & & +~c.c.) & &
+~1_{2,2,0}+1_{2,-2,0}+c.c.)
\end{array}
\end{equation}
\paragraph{$G_{p}=E_{8}:$}
\begin{equation}
\small
\begin{array}{c@{}c@{}l@{}c@{}l@{}c@{}l}
E_8&\rightarrow&E_7\times U(1) &\rightarrow&
E_6\times U(1)^2 &\rightarrow& SO(10)\times U(1)^2\times U(1)_S\\
248 &\rightarrow & ~133_0+1_0 & \rightarrow & ~78_{0,0}+2\times
1_{0,0} &\rightarrow &
~45_{0,0,0}+3\times 1_{0,0,0}+16_{0,0,-3}+\overline{16}_{0,0,3}\\
& & +{\it 1_{\pm2}} & & +27_{0,2}+\overline{27}_{0,-2}+{\it
1_{\pm2,0}}
& & +(16_{0,2,1}+10_{0,2,-2}+1_{0,2,4}+c.c.)+{\it 1_{\pm2,0,0}} \\
& & +56_1 & & +27_{1,-1}+\overline{27}_{1,1}+1_{1,\pm3}&
&+ 16_{1,-1,1}+10_{1,-1,-2}+1_{1,-1,4} \\
& & & & & &
+\overline{16}_{1,1,-1}+\overline{10}_{1,1,2}+1_{1,1,-4}+
1_{1,\pm3,0}  \\
& & +{\it {56}_{-1}} & & +{\it
27_{-1,-1}+\overline{27}_{-1,1}+1_{-1,\pm3}} & &
+ \it 16_{-1,-1,1}+10_{-1,-1,-2}+1_{-1,-1,4}\\
& & & & & & +\it
\overline{16}_{-1,1,-1}+\overline{10}_{-1,1,2}+1_{-1,1,-4}+
1_{-1,\pm3,0}
\end{array} \label{e6161610}
\end{equation}
\begin{equation}
\small
\begin{array}{c@{}c@{}l@{}c@{}l@{}c@{}l}
E_8&\rightarrow&E_7\times U(1) &\rightarrow&
SO(12)\times U(1)^2 &\rightarrow& SO(10)\times U(1)^2\times U(1)_S\\
248 &\rightarrow & ~133_0+1_0 & \rightarrow & ~66_{00}+2\times
1_{00}+{\it 1_{0,\pm2}} &\rightarrow &
~45_{0,0,0}+3\times 1_{0,0,0}+(10_{0,0,2}+c.c.)+{\it 1_{0,\pm2,0}}\\
& & +{\it 1_{\pm2}} & & +32'_{0,1}+{\it{32}'_{0,-1}}+{\it
1_{\pm2,0}} & &
+16_{0,1,-1}+\overline{16}_{0,1,1}+{\it 16_{0,-1,-1}+\overline{16}_{0,-1,1}}+{\it 1_{\pm2,0,0}} \\
& & +56_1 & & +32_{1,0}+12_{1,1}+{\it{12}_{1,-1}}&
&+ 16_{1,0,1}+\overline{16}_{1,0,-1}+10_{1,1,0}+1_{1,1,\pm2} \\
& & & & & & +{\it {10}_{1,-1,0}+1_{1,-1,\pm2}}\\
& & +{\it{56}_{-1}}& & +32_{-1,0}+12_{-1,1}+{\it{12}_{-1,-1}} & &
+ 16_{-1,0,1}+\overline{16}_{-1,0,-1}+10_{-1,1,0}+1_{-1,1,\pm2} \\
& & & & & & +{\it {10}_{-1,-1,0}+1_{-1,-1,\pm2}}
\end{array}  \label{so12161610}
\end{equation}

\subsubsection{$G_{S}=SO(12)$}

For $G_{S}=SO(12)$, we have following enhancement patterns
\[SO(12)\rightarrow SO(14)\rightarrow SO(16)\]
with $G_{p}=SO(16)$, and
\[SO(12)\rightarrow E_{7}\rightarrow E_{8}\]
with $G_{p}=E_{8}$.

In this case, at $G_{p}=SO(16)$, we have couplings ${\bf
10}\,\overline{\bf 10}\, \bf 1$ and $\bf 10\,10\,1$, and at
$G_{p}=E_{8}$, we can obtain $\bf 16\,16\,10$.

\subsubsection{$G_{S}=E_{6}$}

For $G_{S}=E_6$, we have the following enhancement pattern
\[E_{6}\rightarrow E_{7}\rightarrow E_{8}\]
with $G_{p}=E_{8}$.

In this case, the only $G_{p}$ we get is $E_{8}$, which gives rise
to the couplings $\bf 16\,16\,10$.


\end{document}